\documentclass[twocolumn,10pt,cleanfoot]{asme2e}

\usepackage{epsfig} 
\DeclareMathOperator*{\argmax}{arg\,max}

\title{Multi-Information Source Fusion and Optimization to Realize ICME: Application to Dual Phase Materials}

\author{Seyede Fatemeh Ghoreishi, 
    \affiliation{
    Graduate Research Assistant\\
	Dept. of Mechanical Eng.\\
	Texas A\&M University\\
	College Station, TX 77843\\
    Email: f.ghoreishi88@tamu.edu
    }	
}

\author{Abhilash Molkeri
         \affiliation{ Graduate Research Assistant\\
         Dept. of Materials Science and Eng.\\
        	Texas A\&M University\\
	College Station, TX 77843\\
    Email: abhilashmolkeri@tamu.edu
        }
}

\author{Ankit Srivastava
         \affiliation{ Assistant Professor\\
         Dept. of Materials Science and Eng. \\
        	Texas A\&M University\\
	College Station, TX 77843\\
    Email: ankit.sri@tamu.edu
        }
}

\author{Raymundo Arroyave
         \affiliation{ Professor\\
         Dept. of Materials Science and Eng.\\
        	Texas A\&M University\\
	College Station, TX 77843\\
    Email: rarroyave@tamu.edu
        }
}

\author{Douglas Allaire\thanks{Address all correspondence to this author.} 
    \affiliation{ Assistant Professor\\
	Dept. of Mechanical Eng.\\
	Texas A\&M University\\
	College Station, TX 77843\\
    Email: dallaire@tamu.edu
    }
}

\begin{document}

\maketitle    

\begin{abstract}
{\it Integrated Computational Materials Engineering (ICME) calls for the integration of computational tools into the materials and parts development cycle, while the Materials Genome Initiative (MGI) calls for the acceleration of the materials development cycle through the combination of experiments, simulation, and data. As they stand, both ICME and MGI do not prescribe how to achieve the necessary tool integration or how to efficiently exploit the computational tools, in combination with experiments, to accelerate the development of new materials and materials systems. This paper addresses the first issue by putting forward a framework for the fusion of information that exploits correlations among sources/models and between the sources and `ground truth'. The second issue is addressed through a multi-information source optimization framework that identifies, given current knowledge, the next best information source to query and where in the input space to query it via a novel value-gradient policy.  The querying decision takes into account the ability to learn correlations between information sources, the resource cost of querying an information source, and what a query is expected to provide in terms of improvement over the current state.  The framework is demonstrated on the optimization of a dual-phase steel to maximize its strength-normalized strain hardening rate. The ground truth is represented by a microstructure-based finite element model while three low fidelity information sources---i.e. reduced order models---based on different homogenization assumptions---isostrain, isostress and isowork---are used to efficiently and optimally query the materials design space.  }
\end{abstract}

%


\section{Introduction}
\label{sec:intro}
\subsection{Motivation: Towards Accelerated Materials Design}
\label{subsec:motivation}
Over the past two decades, there has been considerable interest in the development of frameworks to accelerate the materials development cycle. Back in the late 90s, Greg Olson popularized the concept of materials-as-hierarchical-systems~\cite{olson1997computational,olson2000designing}, amenable for improvement through the exploitation of \emph{explicit}  processing-structure-properties-performance (PSPP) relationships. Olson used this framework to develop (inverse) linkages connecting performance/property requirements to desired (multi-scale) structural features and the latter to the corresponding processing steps. A decade later, the Integrated Computational Materials Engineering (ICME)~\cite{allison2011integrated,national2008integrated} framework prescribed the combination of theory, experiments and computational tools to streamline and accelerate the materials and manufacturing development cycle. Similarly, the Materials Genome Initiative~\cite{holdren2011materials} calls for the acceleration of the materials development cycle through the combination of experiments, simulations and data. We would like to point out that ICME and MGI are \emph{aspirational} in that the former does not prescribe the way to carry out the \emph{integration} of multiple tools and the latter does not put forward a feasible strategy to \emph{accelerate} the materials development cycle. On the other hand, ICME and MGI have motivated considerable development in terms of the sophistication in the tool sets used to carry out the computer-assisted exploration of the materials design space~\cite{madison2016integrated,agrawal2016perspective,kalidindi2015materials,voorhees2015modeling}. 

\subsection{Challenges and Opportunities}
\label{subsec:challenge}
On the integration front, it has long been recognized that in order to establish quantitative PSPP relationships it is necessary to integrate multiple (computational) tools across multiple scales~\cite{voorhees2015modeling}. Realizing such integration is a necessary (albeit, not sufficient) condition to achieving any measure of success when attempting to carry out computationally-assisted materials development exercises. Explicit integration of multiple tools is technically challenging, particularly because of the considerable expense of computational models, the complexity of the input/output interfaces of such models, and the asynchronous nature of the development of such tools. We would like to note, however, that some efforts have recently emerged that attempt to explicitly integrate models within a single framework for materials design~\cite{reddy2017ontological,savic2017integrated,diehl2017identifying}.  Approaches that instead use statistical techniques and machine learning tools to better sample the design space have proven to be effective~\cite{bessa2017framework}.

Another strategy for the accelerated discovery of materials (most closely associated to the MGI) has been the use of high-throughput (HT) experimental~\cite{potyrailo2011combinatorial,suram2015generating,green2017fulfilling} and computational~\cite{curtarolo2013high} approaches that, while powerful, have important limitations as they tend to be sub-optimal in resource allocation as experimental decisions do not account for the cost and time of experimentation. Resource limitation cannot be overlooked as it is often the case that once a bottle-neck in HT workflows has been eliminated (e.g., synthesis of ever more expansive materials libraries), another one suddenly becomes apparent (e.g., need for high-resolution characterization of materials libraries). 

Recently, notions of optimal experimental design, within the overall framework of Bayesian Optimization (BO), have been put forward as a strategy to overcome the limitations of traditional (costly) exploration of the design space. For example, Balachandran et al.~\cite{balachandran2016adaptive} have put forward a framework that balances the need to exploit current knowledge of the design domain with the need to explore it by using a metric that evaluates the value of the next experiment (or simulation) to carry out. BO-based approaches rely on the construction of a response surface of the design space and are typically limited to the use of \emph{a single model} to carry out the queries. This is an important limitation, as often times, at the beginning of a materials discovery problem, there is not sufficient information to elucidate the feature set (i.e., model) that is most related to the specific performance metric to optimize. 

Talapatra et al.~\cite{talpatra2018towards} recently proposed a framework that is capable of adaptively selecting competing models connecting materials features to performance metrics through Bayesian Model Averaging (BMA), followed by optimal experimental design.
Ling et al.~\cite{ling2006managing} propose a value of information framework that is capable of managing information from multiple sources with a particular emphasis on imprecise probabilities. 
Also, there has been recent work on non-hierarchical fusion for design that has led to promising avenues for information source integration~\cite{chen2016multimodel,lam2015multifidelity,allaire2014mathematical} that we generally build off here.  

\subsection{Description of this Work}
\label{subsec:description}
It is clear from the brief discussion above that, while considerable progress has been made recently in the development of novel frameworks for accelerating materials development efforts, several important challenges remain to be solved. Model-based ICME-style frameworks tend to focus on integrating tools at multiple levels under the assumption that there is \emph{a single} model/tool relevant at a specific scale of the problem. This precludes the use of multiple models that may be more/less effective in different regions of the performance space. Data-centric approaches, on the other hand, tend to focus (with some exceptions) on the brute-force exploration of the materials design space, without accounting for the considerable cost associated with such exploration.

In this work, we present a framework that addresses the two outstanding issues listed above in the context of the optimal microstructural design of ductile multi-phase materials, such as advanced high strength steels. Specifically, we carry out the fusion of multiple information sources that connect microstructural descriptors to mechanical performance metrics.  This fusion is done in a way that accounts for and exploits the correlations between each individual information source---reduced order model constructed under different simplifying assumptions regarding the partitioning of (total) strain, stress or deformation work among the phases constituting the microstructure---and between each information source and the ground truth---represented in this case by a full-field microstructure-based finite element model. 
We note here that while this finite element model is computational, and thus could be considered as a higher fidelity model as part of a multifidelity framework, our intention is create a framework for predicting ground truth.  Specifically, we are not interested in matching the highest fidelity model, but in predicting material properties when created at ground truth.  There is usually no common resource tradeoff in this scenario, which is in contrast to traditional computational multifidelity frameworks that trade computational expense and accuracy.  Thus, the finite element model is used here as a proxy for a ground truth experiment, and is treated as such in the demonstrations provided.

In our framework, we value the impact a new query to an information source has on the fused model.  In particular, we perform the search over the input domain and the information source options concurrently to determine which next query will lead to the most improvement in our objective function.  This concurrent approach, to our knowledge, has not been addressed in the literature.  In addition, our exploitation of correlations between the discrepancies of the information sources in the fusion process differs significantly from previous work and enables the identification of ground truth optimal points that are not shared by any individual information sources in the analysis.

The remainder of the paper is as follows: First, we proceed to motivate the work in the context of microstructure-sensitive modeling and design of dual-phase ductile materials, e.g., advanced high strength steels. These advanced structural alloys are one of the most technologically sought after materials used in lightweight structural applications, such as automotive manufacturing. Next, we describe a microstructure-based finite element model---considered in this work as the `ground truth'---for predicting the stress-strain response of ductile dual-phase materials as well as the reduced order models that predict the stress-strain response of multi-phase microstructures under different assumptions regarding the partitioning of stress, strain or work of deformation. We then present and demonstrate the proposed framework for correlation-exploiting information fusion through \emph{reification}. This is followed by the description  and demonstration of the proposed multi-information optimization framework. We close the paper by discussing further directions for the current research program. 


\section{Mechanical Behavior of Dual-Phase Microstructures}
\label{sec:mat}

A class of one of the most technologically sought after structural materials, known as advanced high-strength steels, derive their exceptional properties from complex, heterogeneous microstructures. Of the various advanced high strength steels, dual-phase steels have experienced the fastest growth in the automotive industry \cite{bhattacharya2011metallurgical}. These dual-phase advanced high strength steels primarily consist of hard martensite islands dispersed in a soft ferrite matrix~\cite{rashid1981dual}. Both these phases undergo non-linear elastic-plastic deformation with strikingly different strength levels and strain hardenability~\cite{chen2014microscale, srivastava2016multiscale}. The overall mechanical properties of dual-phase steels are thus determined partly by the mechanical properties of the constituent phases, and partly by the microstructural features, such as the volume fraction of the phases. The properties of the phases and the microstructural features can, in principle, be tuned and optimized to achieve a particular performance matrix. 

The microstructure-property correlation of ductile dual-phase materials can be explored by high-fidelity microstructure-based finite element models. However, these come at considerable computational cost that precludes their use to carry out search in the microstructure space for regions of optimal performance. The response of composite microstructures consisting of more than one phase can be approximated through the use of low-fidelity models based on different assumptions underlying the homogenized response of the multi-phase microstructure. As described below, here we will carry out the fusion of multiple reduced-order models  based on isostrain~\cite{voigt1889relation}, isostress~\cite{reuss1929berechnung} or isowork~\cite{bouaziz2002mechanical} assumptions for the partitioning of the macroscopic strain, stress or work, respectively, among the microstructural constituents. 

In this work, we will exploit statistical correlations between the different information sources to arrive at a fused model with significantly better fidelity with respect to the ground truth (microstructure-based finite element model) than any individual source (reduced-order model). The fused model will then be integrated with a Bayesian sequential design optimization framework to arrive at optimal microstructures that maximize the strength normalized strain-hardening rate by identifying and exploiting optimal sequential queries of different information sources. The quantity of interest, strength normalized strain-hardening rate, is an important manufacturing-related attribute as it dictates the ductility and formability of the material. The details of the microstructure-based finite element modeling (ground truth) of dual-phase microstructures and the three lower fidelity reduced-order models (information sources) are described below.

\subsection{Microstructure-based Finite Element Model}

Microstructure-based finite element modeling is carried out to calculate the overall mechanical response of the ductile dual-phase microstructures. To this end, we generate 3D representative volume elements (RVEs) of the dual-phase microstructures following the procedure detailed in ~\cite{gerbig2017analysis}. The RVE is a composite dual-phase microstructure with two discretely modeled phases: a soft phase representative of the ferrite phase and a hard phase representative of the martensite phase, present in dual-phase advanced high strength steels. A typical 3D RVE of the dual-phase microstructure is shown in the inset of Fig.~\ref{Figure: SS}. The RVE consists of 27,000 C3D8 brick elements of the ABAQUS/standard element library \cite{documentation2010version}, and has a dimension of $100 \mu\textrm{m} \times 100 \mu\textrm{m} \times 100 \mu\textrm{m}$. The volume fraction of the phases in the RVE is always an integral multiple of the volume of one element, which is $3.7 \times 10^{-5} \mu\textrm{m}^3$. The RVE is subjected to fully periodic boundary conditions on all six faces and monotonically increasing uniaxial tensile deformation. This allows for the calculation of the overall uniaxial tensile stress-strain response of the composite microstructure.

In the calculations, both phases are assumed to follow isotropic elastic-plastic stress-strain response. The Young's modulus of both phases is taken to be $E=200$GPa and Poisson's ratio is taken to be $\nu=0.3$. The plastic response of both phases are modeled using the Ludwik power law constitutive relation,

\begin{equation} 
\tau = \tau_{o} + K(\epsilon_{pl})\,^{n},
\label{ludwik}
\end{equation}
where $\tau$ is the flow stress, $\epsilon_{pl}$  is the equivalent plastic strain, $\tau_{o}$  is the yield strength, $K$ is the strengthening coefficient, and $n$ is the strain hardening exponent. The values of $\tau_{o}$, $K$, and $n$ for the constituent phases are given in Tab.~\ref{Tab:1}. The parameters are chosen to represent lower initial yield strength of the ferrite (soft) phase compared to the martensite (hard) phase, and higher strain hardenability of the ferrite phase compared to the martensite phase \cite{srivastava2016multiscale, gerbig2017analysis}.

\begin{table}[!h]
\vskip -.5cm
\caption{Constitutive parameters for the constituent phases}
\begin{center}
\begin{tabular}{c l l l } 
& \\
 \hline
 \textbf{Constituent Phase} & \textbf{$\tau_{o}$ [MPa]} & \textbf{$K$ [MPa]} & \textbf{$n$}\\
 \hline
 Soft (ferrite) & 300 & 2200 & 0.5\\ 
 Hard (martensite) & 1500 & 450 & 0.06\\ 
 \hline
\end{tabular}
 \label{Tab:1}
 \end{center}
 \vskip -.01cm
\end{table}

\subsection{Reduced-order Models}

We use three low-fidelity reduced-order models as three sources of information. These three reduced-order models are:

\begin{itemize}
\item[i.]
The Voigt/Taylor isostrain model, where the basic assumption is that the strain field is uniform among the constituent phases~\cite{nemat2013micromechanics}. The effective stress is expressed in terms of the local stress average with respect to both phases weighted by their respective volume fractions. That is, for this model we have
\begin{equation} 
\epsilon^{T}_{pl}=\epsilon^{h}_{pl}=\epsilon^{s}_{pl}, \quad \tau^{T}=f_{\textrm{hard}} \tau^{h} +(1-f_{\textrm{hard}})\tau^{s}.
\label{iso-strain}
\end{equation}

\item[ii.]
The Reuss/Sachs isostress model, where the basic assumption is that the stresses among the phases are homogeneous~\cite{nemat2013micromechanics}. The effective strain is calculated in terms of the average of the strains in each phase weighted by their respective volume fractions. Thus, for this model we have
\begin{equation} 
\tau^{T}=\tau^{h}=\tau^{s}, \quad \epsilon^{T}_{pl}=f_{\textrm{hard}} \epsilon^{h}_{pl}+(1-f_{\textrm{hard}})\epsilon^{s}_{pl}. 
\label{iso-stress}
\end{equation}

\item[iii.]
The isowork model, which is an approximation based on the principle that work of deformation is equally distributed in all the constituent phases in the dual-phase microstructure at any strain level.  That is,
\begin{equation} 
\tau^{h}\epsilon^{h}_{pl}=\tau^{s}\epsilon^{s}_{pl}.
\label{iso-work}
\end{equation}

\end{itemize}
In Eqs.~\ref{iso-strain}, \ref{iso-stress} and \ref{iso-work}, $\epsilon^{T}_{pl}$ is the overall plastic strain, $\epsilon^{h}_{pl}$ is the plastic strain in the hard (martensite) phase, $\epsilon^{s}_{pl}$ is the plastic strain in the soft (ferrite) phase, $\tau^{T}$ is the overall stress, $\tau^{h}$ is the stress in the hard (martensite) phase, $\tau^{s}$ is the stress in the soft (ferrite) phase, and $f_{\textrm{hard}}$ is the volume fraction of the hard phase in the microstructure. The stress-strain relations, $\tau=f(\epsilon_{pl})$, of both phases are assumed to follow, Eq.~\ref{ludwik}, with the values of the parameters given in Tab.~\ref{Tab:1}.

\subsection{Demonstration of Modeling Capabilities}
\label{sec:modcap}

\begin{figure}[!h] 
\centering
\includegraphics[width=0.48\textwidth]{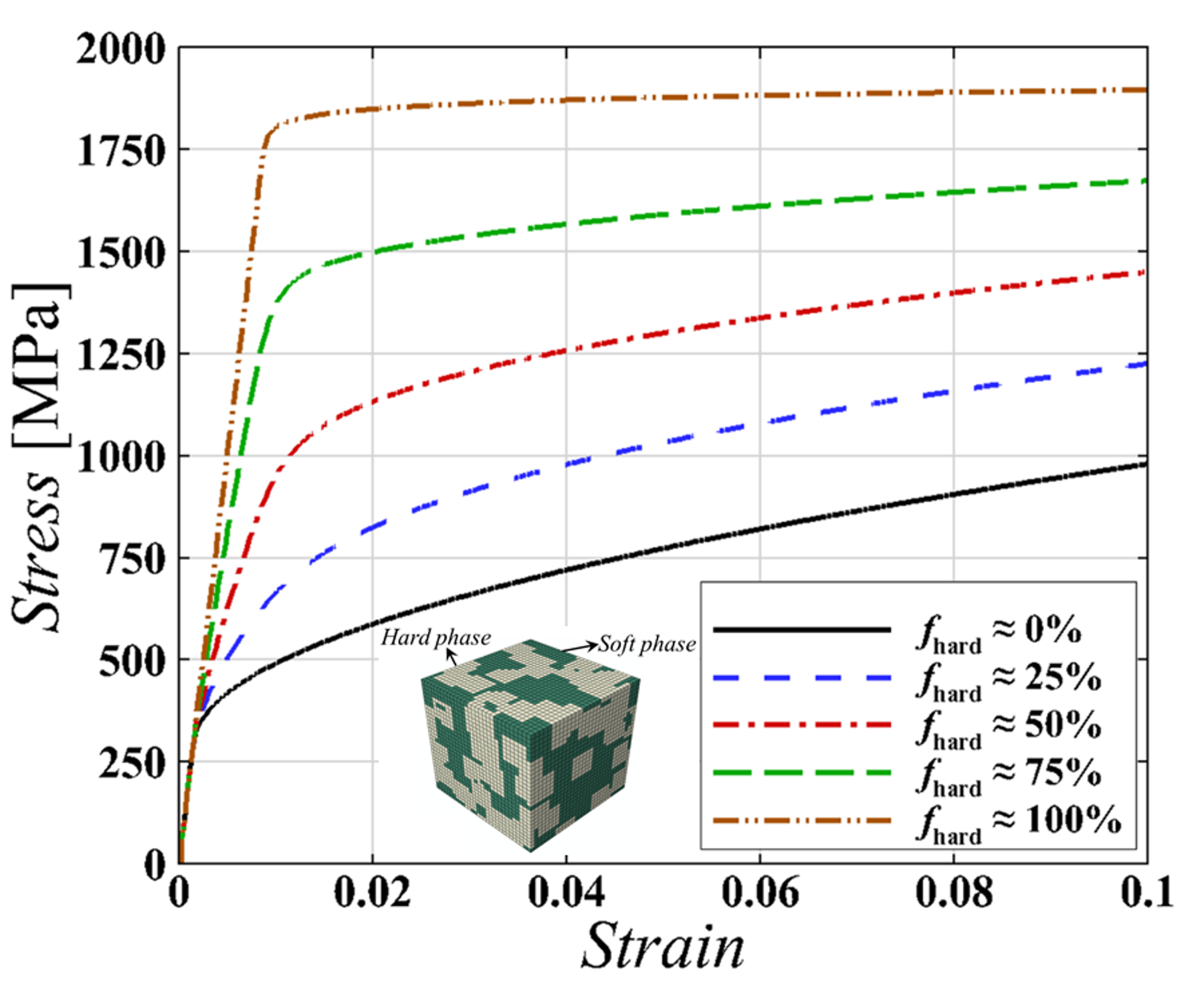}
\caption{The stress-strain response of dual-phase phase microstructures with volume fraction of the hard phase, $f_{\textrm{hard}}=$ $0\%$, $25\%$, $50\%$, $75\%$, and $100\%$. A 3D representative volume element of the dual-phase microstructure is shown in the inset.}
\label{Figure: SS}
\end{figure}

The predicted stress-strain response of dual-phase microstructures with varying volume fraction of the hard phase, $f_{\textrm{hard}}$, using a high-fidelity microstructure-based finite element model is shown in Fig.~\ref{Figure: SS}. As seen in the figure, the flow strength of the dual-phase material increases with increasing volume fraction of the hard phase. But the strain-hardening rate, which is the slope of the stress-strain curve, of the material varies non-monotonically with the volume fraction of the hard phase. 

\begin{figure}[!h] 
\centering
\includegraphics[width=0.48\textwidth]{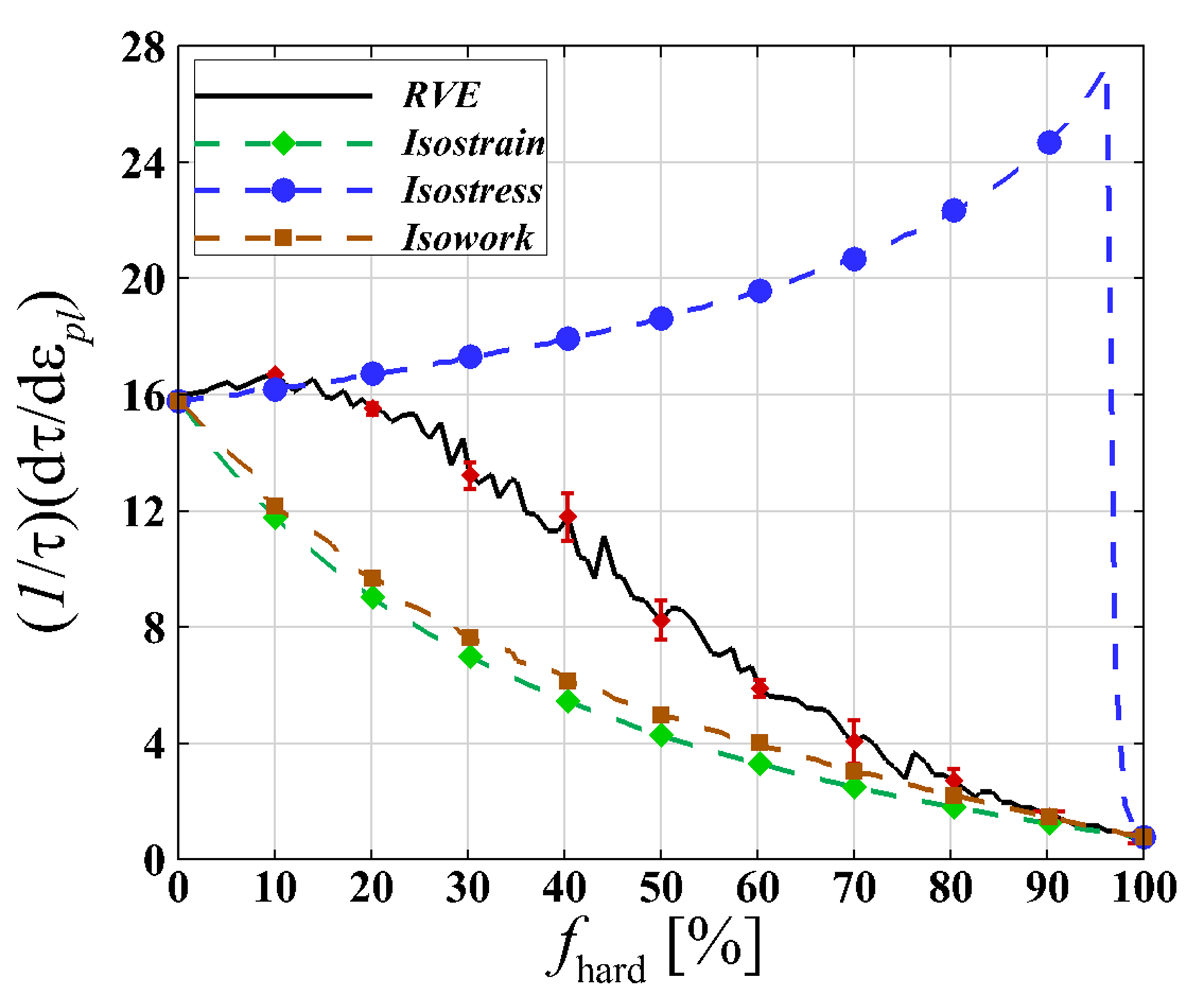}
\caption{Comparison of the variation of the strength normalized strain-hardening rate, $(1/\tau)(\textrm{d}\tau/\textrm{d}\epsilon_{pl})$ at $\epsilon_{pl}=1.5\%$, with the volume fraction of the hard phase, $f_{\textrm{hard}}$, as predicted by the three reduced-order models and the microstructure-based finite element model.}
\label{Figure: hardening}
\end{figure}

The variation of the strength normalized strain-hardening rate, $(1/\tau)(\textrm{d}\tau/\textrm{d}\epsilon_{pl})$, with the volume fraction of the hard phase, $f_{\textrm{hard}}$, estimated at a strain level, $\epsilon_{pl}=1.5\%$, using the microstructure-based finite element calculations, is shown in Fig.~\ref{Figure: hardening}. As seen in the figure, the value of $(1/\tau)(\textrm{d}\tau/\textrm{d}\epsilon_{pl})$ at $\epsilon_{pl}=1.5\%$, first increases with increasing volume fraction of the hard phase and then starts to decrease. In general a higher value of the quantity $(1/\tau)(\textrm{d}\tau/\textrm{d}\epsilon_{pl})$ denotes higher formability of the material. Note, in Fig.~\ref{Figure: hardening}, variation of $(1/\tau)(\textrm{d}\tau/\textrm{d}\epsilon_{pl})$ with $f_{\textrm{hard}}$ exhibits local perturbations. These perturbations are due to the fact that there are several possible realizations of the RVE of a dual-phase microstructure with a fixed volume fraction of the hard phase. These different realizations result in slightly different values of $(1/\tau)(\textrm{d}\tau/\textrm{d}\epsilon_{pl})$ for a fixed $f_{\textrm{hard}}$ value. For a few selected volume fractions of the hard phase, seven realizations of the dual-phase microstructures were generated and their mechanical responses were calculated. The standard error on the values of $(1/\tau)(\textrm{d}\tau/\textrm{d}\epsilon_{pl})$ at $\epsilon_{pl}=1.5\%$ due to different realizations of the dual-phase microstructure with fixed volume fraction of the hard phase are also shown in figure as error bars.

The predictions of $(1/\tau)(\textrm{d}\tau/\textrm{d}\epsilon_{pl})$ at a strain level, $\epsilon_{pl}=1.5\%$, as a function of the volume fraction of the hard phase, $f_{\textrm{hard}}$, using the three low-fidelity reduced-order models are also shown in Fig.~\ref{Figure: hardening}. Compared to the finite element model, the isostress model gives a reasonable prediction of $(1/\tau)(\textrm{d}\tau/\textrm{d}\epsilon_{pl})$ at $\epsilon_{pl}=1.5\%$ for low volume fraction of the hard phase but significantly over-predicts this quantity for large volume fractions of the hard phase. In contrast, the isostrain and isowork models give reasonable predictions at high volume fraction of the hard phase but under-predict $(1/\tau)(\textrm{d}\tau/\textrm{d}\epsilon_{pl})$ at $\epsilon_{pl}=1.5\%$ at lower volume fractions of the hard phase. It is also important to note here that the maximum values of $(1/\tau)(\textrm{d}\tau/\textrm{d}\epsilon_{pl})$ at $\epsilon_{pl}=1.5\%$ according to each information source are significantly different from the ground truth maximum.

\section{Correlation Exploiting Multi-Information Source Optimization} 
\label{sec:Approach}

In most materials design tasks, there are always multiple information sources at the disposal of the designer. For example, the forward connections between microstructures and properties/performance can in principle be developed through experiments as well as (computational) models at different levels of fidelity and resolution. Conventional approaches to ICME, on the other hand, often times make the implicit and unrealistic assumption that \emph{there is only one source available} to query the design space---in this work, our framework uses three relatively simple models (under the isostrain, isostress and isowork approximations) as representative of multiple information sources, while considering a microstructure-sensitive RVE-based simulation as the ground truth.

While information fusion on its own represents a considerable improvement upon the vast majority of ICME-based approaches to materials design currently under development, we posit that an even better approach would necessarily have to account for resource constraints on the exploration of the materials design space (MDS). Specifically, every source used to query the MDS carry a certain (time, monetary, opportunity) cost and thus there are hard limits to the number of queries and the sources used to carry out such queries. Unfortunately, such constraints rarely take a concrete form that can be dealt with using formal constrained optimization approaches.  This is due to the often dynamic nature of the materials design and procurement process.  As a materials design cycle progresses, the current state of the process may dictate if more resources will be allocated to the process or not.  Thus, it is advantageous to tackle such problems in a myopic fashion.  

For single information sources and sequential querying, there are two traditional techniques for choosing what to query next in this myopic context~\cite{scott2011correlated}. These are efficient global optimization (EGO)~\cite{jones1998efficient} and its extensions, such as sequential Kriging optimization (SKO)~\cite{huang2006sequential} and value-based global optimization (VGO)~\cite{moore2014value}, and the knowledge gradient (KG)~\cite{frazier2008knowledge,gupta1994bayesian,gupta1996bayesian}.  EGO uses a Gaussian process~\cite{rasmussen2006gaussian} representation of queried information, but assumes no noise~\cite{schonlau1996global,schonlau1998global}. SKO also uses Gaussian processes, but includes a tunable weighting factor to lean towards decisions with higher uncertainty~\cite{scott2011correlated}.  KG can handle noise and makes its querying selection on the basis of the expected value of the best design after querying.  Here, KG does not require that design to have actually been evaluated by an information source.

Recent developments in~\cite{chen2016multimodel,lam2015multifidelity} extend these sequential optimization approaches to the case of multiple information sources.  The approach we propose here builds off of these approaches by including and exploiting learned correlations in multi-information source fusion and by defining and implementing a two-step lookahead querying strategy referred to as the value-gradient policy.  We describe our formal problem statement, the multi-information source fusion approach, and the value-gradient utility used to guide the querying policy, in the following subsections.

\subsection{Problem Formulation}

A mathematical statement of the problem is formulated as finding the best design, $\mathbf{x}^*$, such that
\begin{equation}
\label{prob}
{\mathbf{x}}^* =  \argmax_{\mathbf{x} \in \chi}  f(\mathbf{x}),
\end{equation}
where $f$ is the ground truth objective function, and $\mathbf{x}$ is a set of design variables in the 
vector space $\chi$.
This ground truth objective function is typically very expensive to query. We note that in this formulation, there is a tacit dynamic constraint on resources.  While ground truth is impractical to query often in an optimization process, other forms of information are usually available and can be used to approximate the ground truth.  These information sources differ in terms of fidelity with respect to the ground truth, as well as resource expenditures required per query. These information sources are also fundamentally related through the fact that they seek to estimate the same quantity of interest.  Thus, there \emph{must} exist statistical correlations between these information sources that can potentially be exploited if learned. In this context, the core issue to myopically addressing Eq.~\ref{prob}, is the decision of what information source to query and where in its input domain to execute that query.  This decision must balance the cost of the query and what that query is expected to tell us about the solution to Eq.~\ref{prob}.

To assign a value to each potential query option over the information source space and the domains of the respective information sources, we create intermediate Gaussian process surrogates for each information source learned from previous queries.  
We assume that we have $S$ information sources, $\bar{f}_i(\mathbf{x})$, where $i \in \{1,2,\ldots,S\}$, available to estimate the ground truth, $f(\mathbf{x})$, at design point $\mathbf{x}$. We further assume that we have $N_i$ previous query results available for information source $i$.  These results are denoted by $\{\mathbf{X}_{N_i}, \mathbf{y}_{N_i}\}$, where $\mathbf{X}_{N_i} = (\mathbf{x}_{1,i}, \ldots, \mathbf{x}_{N_i,i})$ represents the $N_i$ input samples to information source $i$ and $\mathbf{y}_{N_i}$ represents the corresponding outputs from information source $i$. Posterior Gaussian process distributions of each $\bar{f}_i$, denoted as $f_{GP,i}(\mathbf{x})$, at any point $\mathbf{x}$ in the input space, are then given as
\begin{equation}
\label{GP11}
f_{GP,i}(\mathbf{x}) \mid \mathbf{X}_{N_i}, \mathbf{y}_{N_i} \sim \mathcal{N}\left(\mu_i(\mathbf{x}),\sigma_{GP,i}^2(\mathbf{x})\right),
\end{equation}
where
\begin{equation}
\label{mean1}
\mu_i(\mathbf{x}) = K_i(\mathbf{X}_{N_i},\mathbf{x})^T[K_i(\mathbf{X}_{N_i},\mathbf{X}_{N_i})+\sigma^2_{n,i}I]^{-1} \mathbf{y}_{N_i},
\end{equation}
and
\begin{equation}
\begin{aligned}
\label{cov1}
&\sigma_{GP,i}^2 (\mathbf{x}) = \\ &k_i(\mathbf{x},\mathbf{x}) - K_i(\mathbf{X}_{N_i},\mathbf{x})^T[K_i(\mathbf{X}_{N_i},\mathbf{X}_{N_i})+\sigma^2_{n,i}I]^{-1}
K_i(\mathbf{X}_{N_i},\mathbf{x}).
\end{aligned}
\end{equation}
Here, $k_i$ is a real-valued kernel function associated with information source $i$ over the input space, $K_i(\mathbf{X}_{N_i},\mathbf{X}_{N_i})$ is the $N_i \times N_i$ matrix whose $m,n$ entry is $k_i(\mathbf{x}_{m,i}, \mathbf{x}_{n,i})$, $K_i(\mathbf{X}_{N_i}, \mathbf{x})$ is the $N_i \times 1$ vector whose $m^{th}$ entry is $k_i(\mathbf{x}_{m,i}, \mathbf{x})$ for information source $i$, and the term $\sigma^2_{n,i}$ can be used to model observation error for information sources or to guard against numerical ill-conditioning. For the kernel function, we use the commonly used squared exponential kernel function given as
\begin{equation}
\label{eq:5}
k_i(\mathbf{x},\mathbf{x'}) = \sigma_s^2 \exp\left(- \sum_{h = 1}^{d} \frac {(x_h-x'_h)^2}{2l_h^2}\right ),
\end{equation}
where $d$ is the dimension of the input space, $\sigma_s^2$ is the signal variance, and $l_h$, where $h = 1,2,\ldots,d$, is the characteristic length-scale that indicates the correlation between the points within dimension $h$. The parameters of the Gaussian process ($\sigma_s^2$, $l_h$ and $\sigma_n^2$) associated with each information source can be estimated via maximum likelihood or Bayesian techniques~\cite{rasmussen2006gaussian}.

\begin{figure}[h!]
\begin{center}
\includegraphics[width=.45\textwidth]{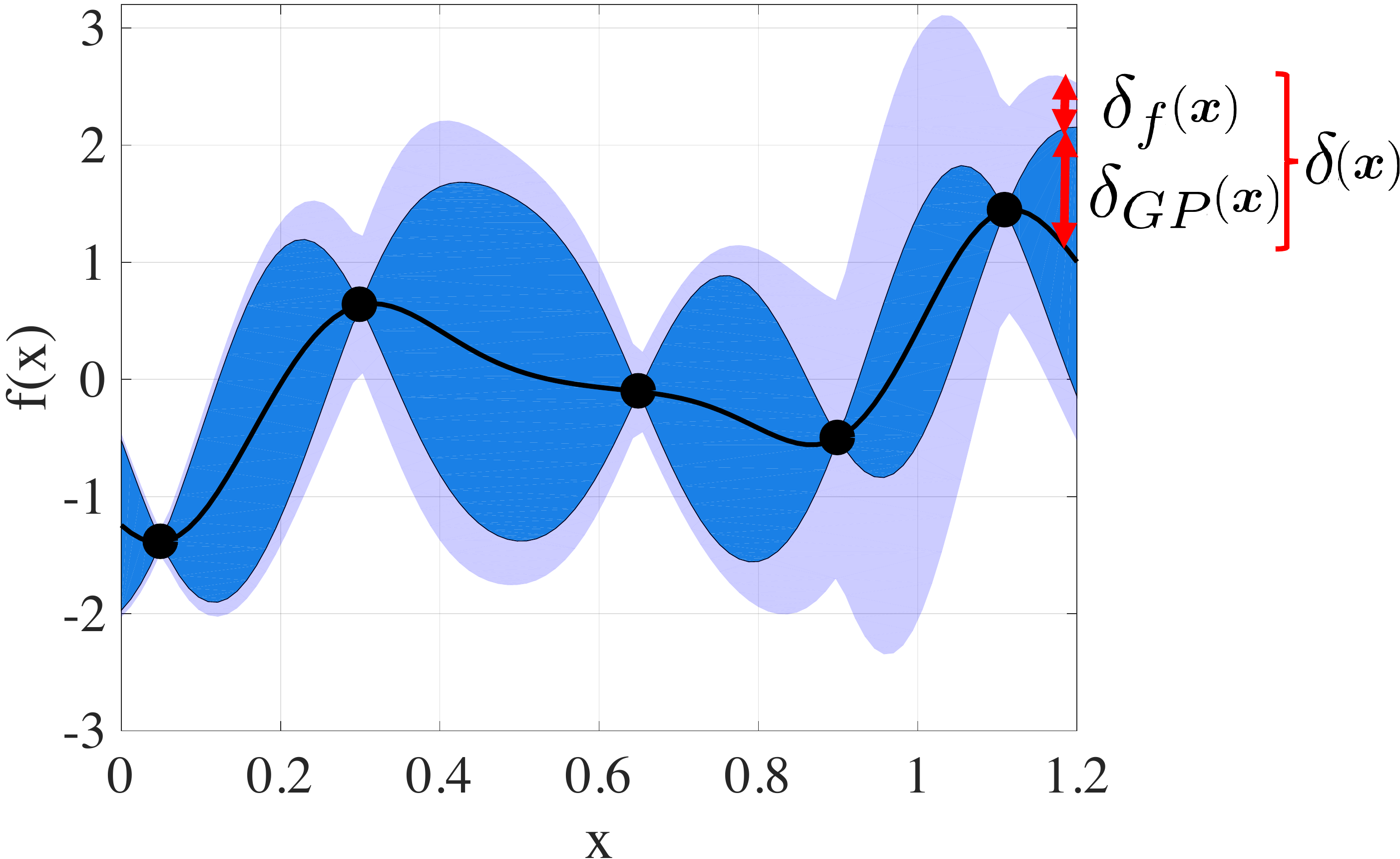}
\end{center}
\caption{A depiction of total uncertainty, which includes both the uncertainty associated with the Gaussian process and uncertainty associated with the fidelity of the information source.}
\label{TotalUncer} 
\end{figure}

To these Gaussian process surrogates, we further quantify the discrepancy of each information source with respect to ground truth. These discrepancies can be estimated from, for example, expert opinion or available ground truth data, and can vary over the input space. We add the estimated uncertainty due to information source discrepancy, $\delta_{f,i}(\mathbf{x})$, to the uncertainty associated with the Gaussian process of information source $i$, denoted by $\delta_{GP,i}(\mathbf{x})$. Specifically, each of the $S$ available information sources for estimating the ground truth objective can be written as
\begin{equation}
\label{IS1mu}
f_i(\mathbf{x}) = \mu_i(\mathbf{x}) + \delta_{i}(\mathbf{x}),
\end{equation}
where 
\begin{equation}
\label{IS1var}
\delta_{i}(\mathbf{x}) = \delta_{GP,i}(\mathbf{x}) + \delta_{f,i}(\mathbf{x}).
\end{equation}
Figure~\ref{TotalUncer} shows a depiction of total uncertainty for an information source, which includes both the uncertainty associated with the Gaussian process and the uncertainty associated with the fidelity of the information source.

\subsection{Correlation Exploiting Fusion}

Available information sources for estimating a ground truth quantity of interest are necessarily correlated by virtue of their estimation task.  If they were not correlated, then presumably they are irrelevant to the estimation task at hand.  
Working under the hypothesis that each information source brings to bear some useful information regarding a quantity of interest, we seek to systematically fuse available information from each source.  Unlike traditional multifidelity methods~\cite{alexandrov2,alexandrov3,Balabanov1998,Balabanov2004,qian2008bayesian,Choi2009,Eldred2004,March2012a,March2012b}, the multi-information source fusion method employed here does not assume a hierarchy of information sources
with the goal of efficiently approximating the highest fidelity source. The goal here is to best approximate a ground truth quantity of interest by leveraging all available information.  This is achieved by learning the correlations between the discrepancies of the available information sources, which results in the ability to mitigate information source bias and avoid overconfidence that arises from the reuse of dependent information.

There are many techniques in use for fusing information from multiple sources of information.  Among these are approaches such as Bayesian modeling averaging~\cite{Leamer_1978, Madigan1994, Draper1995, Hoeting1999, Clyde2003, Clyde2004}, the use of adjustment factors~\cite{Mosleh1986, Zio_1996, Reinert_2006, Riley2011}, covariance intersection methods~\cite{Julier1997, Julier2001}, and fusion under known correlation~\cite{Geisser1965, Morris1977, Winkler_1981}.  
There are also techniques designed to value the improvement potential of a given information source through model refinement~\cite{panchal2008value} and model refinement and selection~\cite{messer2010model}.  In this work, we consider each information source as fixed.  That is, we do not consider improving predictive capabilities of any individual information source and instead focus on leveraging multiple available information sources to construct an improved fused predictive capability.

As noted previously, we hypothesize that every information source contains useful information regarding the ground truth quantity of interest. Thus, as more information sources are incorporated into a fusion process, we expect the variance of quantity of interest estimates to decrease.  This is not necessarily the case for techniques such as Bayesian model averaging and adjustment factors approaches.  For the case of unknown correlations between information sources, recourse must be made to conservative methods, such as covariance intersection.  This method fuses information by assuming the worst case correlation information.  Thus, there is much to gain from estimating correlation between information sources and incorporating these learned correlations in the fusion process.

Since our information sources are represented by intermediate Gaussian processes, their fusion follows that of normally distributed information.
Under the case of known correlations between the discrepancies of information sources, the fused mean and variance are shown to be~\cite{Winkler_1981}
\begin{equation}
\label{e:WinklerMeanGen}
\mathbb{E} [\hat{f}(\mathbf{x})]=\frac{\mathbf{e}^T \tilde{\Sigma}(\mathbf{x})^{-1}  \bm{\mu}(\mathbf{x})}{\mathbf{e}^T \tilde{\Sigma}(\mathbf{x})^{-1} \mathbf{e}},
\end{equation}
\begin{equation}
\label{e:WinklerVarianceGen}
\textrm{Var}\left(\hat{f}(\mathbf{x})\right) = \frac{1}{\mathbf{e}^T \tilde{\Sigma}(\mathbf{x})^{-1} \mathbf{e}},
\end{equation} 
where $\mathbf{e} = [1,\ldots,1]^T$, $\bm{\mu}(\mathbf{x}) = [\mu_1(\mathbf{x}),\ldots,\mu_S(\mathbf{x})]^T$ given $S$ models, and $\tilde{\Sigma}(\mathbf{x})^{-1}$ is the inverse of the covariance matrix between the information sources.  We stress here that the mean as estimated by Eq.~\ref{e:WinklerMeanGen} is not necessarily a convex combination of the information source estimates. For example, in the case of two information sources, Eq.~\ref{e:WinklerMeanGen} is
\begin{equation}
\mathbb{E}[\hat{f}(\mathbf{x})] = \frac{(\sigma_2^2 - \rho \sigma_1 \sigma_2)\mu_1 + (\sigma_1^2 - \rho \sigma_1 \sigma_2) \mu_2}{\sigma_1^2 + \sigma_2^2 - 2\rho \sigma_1 \sigma_2},
\end{equation}
where the dependence on $\mathbf{x}$ has been omitted for notational clarity. If $\sigma_1 < \sigma_2$, then $\mu_1$ will receive a positive weight.  If also, $\rho > \sigma_1/\sigma_2$, then $\mu_2$ will receive a negative weight.  Following Ref.~\cite{Winkler_1981}, a high correlation makes it likely that the estimates will be biased on the same side of the quantity being estimated.  Since the less precise estimate is expected to be further from the true quantity than the more precise estimate, the less precise estimate receives a negative weight.  This acts to shrink the fused estimate towards the true quantity of interest.  This enables the mean of the fused estimate to be outside the bounds of the means of any of the individual information source estimate means. 

The key to the proper use of fusion of normally distributed information is the estimation of the correlation coefficients over the domain.  For this, we use the reification process defined in~\cite{allaire2012fusing, thomison2017model}. In this process, to estimate the correlation coefficients between the deviations of information sources $i$ and $j$, each of the information sources $i$ and $j$, one at a time, is {\em reified}, or treated as a ground truth model. Assuming that information source $i$ is reified, 
the correlation coefficients between the information sources $i$ and $j$, for $j = 1, \dots, i-1, i+1, \dots, S$, are given as
\begin{equation}
\rho_{ij}(\mathbf{x}) = \frac{\sigma_i^2(\mathbf{x})}{\sigma_i(\mathbf{x}) \sigma_j(\mathbf{x})} = \frac{\sigma_i(\mathbf{x})}{\sqrt{\left(\mu_i(\mathbf{x})-\mu_j(\mathbf{x})\right)^2 + \sigma_i^2(\mathbf{x})}},
\label{e:rho1}
\end{equation}
where $\mu_i(\mathbf{x})$ and $\mu_j(\mathbf{x})$ are the mean values of the Gaussian processes of information sources $i$ and $j$ respectively, at design point $\mathbf{x}$, and $\sigma_i^2(\mathbf{x})$ and $\sigma_j^2(\mathbf{x})$ are the total variances at point $\mathbf{x}$. Afterward, information source $j$ is reified to estimate $\rho_{ji}(\mathbf{x})$. Then, the variance weighted average of the two estimated correlation coefficients is used as the estimate of the correlation between the errors as
\begin{equation}
\bar{\rho}_{ij}(\mathbf{x}) = \frac{\sigma_j^2(\mathbf{x})}{\sigma_i^2(\mathbf{x}) + \sigma_j^2(\mathbf{x})} \rho_{ij}(\mathbf{x}) + \frac{\sigma_i^2(\mathbf{x})}{\sigma_i^2(\mathbf{x})+\sigma_j^2(\mathbf{x})} \rho_{ji}(\mathbf{x}).
\label{e:VWrho}
\end{equation}
These average correlations are then used 
to estimate the fused mean and variance in Eqs.~\ref{e:WinklerMeanGen} and \ref{e:WinklerVarianceGen}.

\subsection{Value-Gradient Querying}

\begin{figure}[h!]
\begin{center}
\includegraphics[width=.5\textwidth]{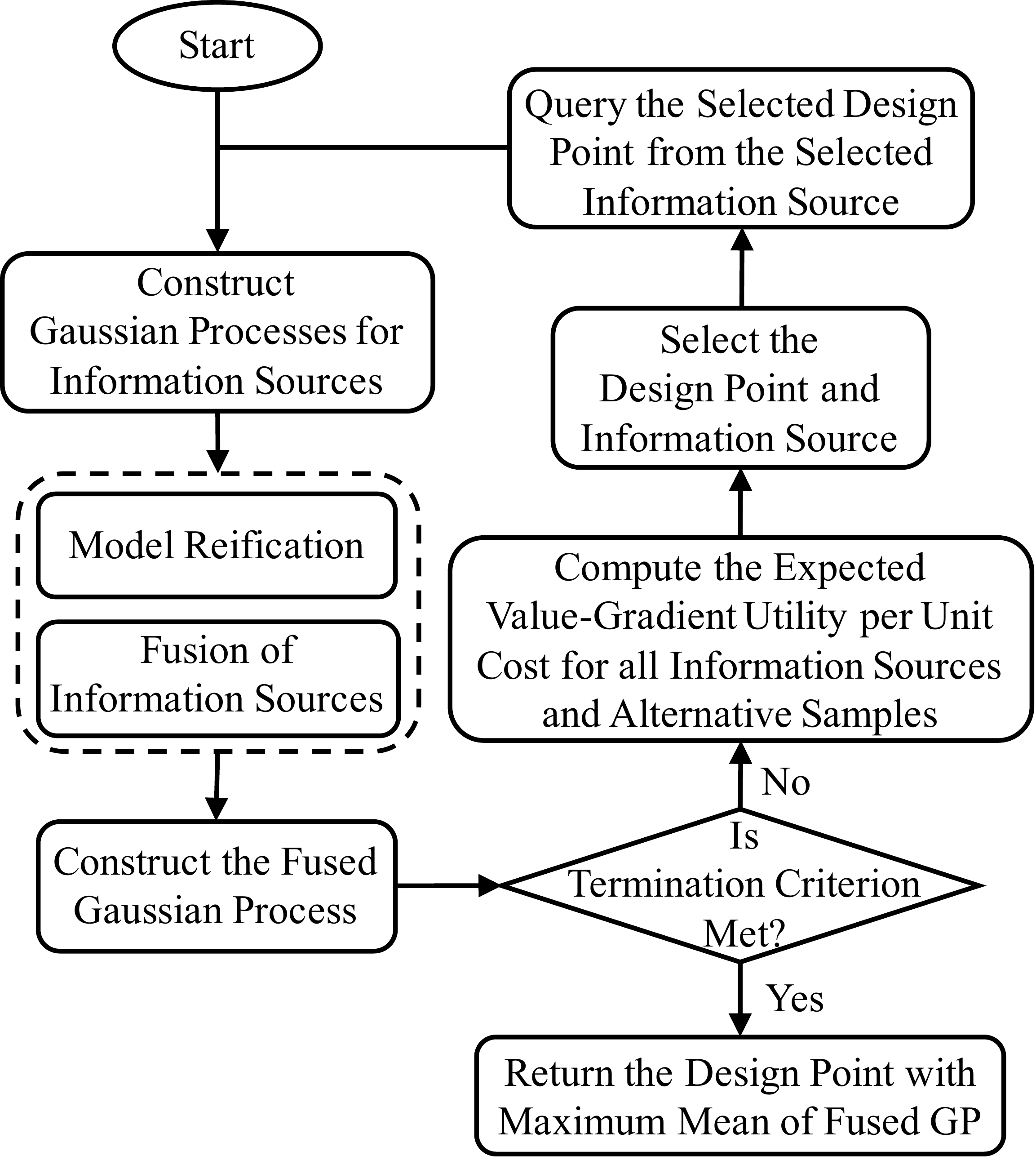}
\end{center}
\caption{Flowchart of the proposed approach.}
\label{FlowChart} 
\end{figure}

By computing the fused means and variances in the input design space $\chi$, we construct a fused Gaussian process over this space.  This fused information source contains all of our current knowledge about the ground truth objective function.
Our goal is to optimize ground truth by leveraging new queries to the less resource-expensive information sources.  Once our resources for information source querying have been exhausted, the predicted ground truth optimal design can be synthesized or produced in an experiment.  This information can of course then be fed back to the multi-information source optimization framework and used to update information source discrepancy and correlation information.
A flowchart of our proposed framework is presented in Fig.~\ref{FlowChart}.  We describe the value-gradient utility in the following paragraphs.

The task then, is to determine what information source to query and where to query it, concurrently, so as to produce the most value in terms of addressing Eq.~\ref{prob}, with the tacit resource constraint in mind.
For this decision, we propose a utility, which we refer to as the value-gradient utility, which takes into account both the immediate improvement in one step and expected improvement in two steps. The idea here being that we seek to produce rapid improvement, with the knowledge that every resource expenditure could be the last, but we also seek to position ourselves best for the next resource expenditure.  In this sense, we are equally weighting next step value with next step (knowledge) gradient information, hence the term value-gradient.

The immediate improvement can be quantified by the maximum mean function value of the fused Gaussian process, $\mu^*_{\textrm{fused}}$. Since the best estimate of the objective function is represented by 
the fused Gaussian process, which is the probabilistic representation of the ground truth objective function,
there is uncertainty in the value of the predicted ground truth objective function upon querying the next sample.
Thus, we compute the expected value of improvement using the posterior predictive distribution of the fused model.
Letting $(i_{1:N},\mathbf{x}_{1:N}, y_{1:N})$ be the information sources, design points, and the corresponding objective values used for the first $N$ queries and $\hat{f}$ denote the posterior distribution of the fused model, the expected improvement ($EI$) at design point $\mathbf{x}$ is defined as
\begin{equation}
\begin{aligned}
EI(\mathbf{x}) &= \mathbb{E} \left[ \max_{\mathbf{x}' \in \chi} \mathbb{E}[\hat{f}(\mathbf{x}') \mid i_{1:N}, \mathbf{x}_{1:N}, \mathbf{x}_{N+1}= \mathbf{x},y_{1:N}]\right.\\
&\,\,\,\,\quad\qquad\qquad\left.-\max_{\mathbf{x}' \in \chi} \mathbb{E}[\hat{f}(\mathbf{x}') \mid i_{1:N}, \mathbf{x}_{1:N}, y_{1:N}] \right]\\
&= \mathbb{E} \left[ \max_{\mathbf{x}' \in \chi} \mathbb{E}[\hat{f}(\mathbf{x}') \mid i_{1:N}, \mathbf{x}_{1:N}, \mathbf{x}_{N+1}= \mathbf{x},y_{1:N}] \right]\\
&\,\,\,\,\,\quad\qquad\qquad-\max_{\mathbf{x}' \in \chi} \mathbb{E}[\hat{f}(\mathbf{x}') \mid i_{1:N}, \mathbf{x}_{1:N}, y_{1:N}],
\end{aligned}
\end{equation}
where the last expression comes out of the expectation operator as it is a known value when conditioned on the first $N$ queries.

The KG policy of~\cite{frazier2008knowledge, powell2012optimal, frazier2009knowledge} takes an information-economic approach to maximize this expectation. Letting $H^N = \mathbb{E}[\hat{f}(\mathbf{x})\mid i_{1:N}, \mathbf{x}_{1:N}, y_{1:N}]$ be the knowledge state, the value of being at state $H^N$ is defined as $V^N(H^N)=\max\limits_{\mathbf{x} \in \chi} H^N$. The knowledge gradient, which is a measure of expected improvement, is defined as
\begin{equation}
\nu^{KG}(\mathbf{x})=\mathbb{E}[V^{N+1}\left(H^{N+1}(\mathbf{x})\right)-V^{N}(H^{N}) \mid H^N].
\label{knowledgegradient}
\end{equation}
The KG policy for sequentially choosing the next query is then given as
\begin{equation}
\mathbf{x}^{KG} = \arg\max_{\mathbf{x} \in \chi}  \nu^{KG}(\mathbf{x}).
\label{maxKG}
\end{equation}
Calculation of the knowledge gradient is discussed in detail in two algorithms presented in~\cite{frazier2009knowledge}.  The method has been shown in Refs.~\cite{frazier2008knowledge,frazier2009knowledge} to perform very well when faced with highly nonlinear and multimodal objective functions.  

Given both immediate and expected improvement, our proposed value-gradient utility is given as
\begin{equation}
\label{utility}
U = \mu^*_{\textrm{fused}} + \max_{\mathbf{x} \in \chi } \;  \nu^{KG}(\mathbf{x}),
\end{equation}
where the first term is the maximum value of the mean function of the current fused model and the second term is the maximum expected improvement that can be obtained with another query as measured by the knowledge gradient over the fused model.  We can then define a value-gradient policy as the policy that selects the next query such that the value-gradient utility is maximized. By considering the immediate gain in the next step and the expected gain in the step that follows, the value-gradient is a two-step look-ahead policy.

To determine the next design point and information source to query efficiently, we generate Latin hypercube samples in the input design space as alternatives denoted as $\mathbf{X}_f$.  For low dimensional problems, a uniform grid of alternatives could also be considered. Among these alternatives, we seek to find the query that maximizes the value-gradient utility of Eq.~\ref{utility}.
According to Eq.~\ref{GP11}, an evaluation of information source $i$, at design point $\mathbf{x}$, is distributed normally with mean $\mu_{i}(\mathbf{x})$ and variance $\sigma^2_{GP,i}(\mathbf{x})$. For a given alternative, $\mathbf{x}$, we draw $N_q$ independent samples from the distribution at that point as
\begin{equation}
\label{samps}
f_i^q(\mathbf{x}) \sim \mathcal{N}\left(\mu_i(\mathbf{x}),\sigma_{GP,i}^2(\mathbf{x})\right), \,\, i = 1,\dots, S\,\, \text{and}\,\, q = 1, \dots, N_q.
\end{equation}
In order to predict the impact of querying each alternative on the utility function, we temporarily augment the design point, $\mathbf{x}$, and the sampled information source output value, $f_i^q(\mathbf{x})$, one at a time, to the available samples of information source $i$. By adding this sample, the Gaussian process of information source $i$ and as a result, the fused Gaussian process, are temporarily updated. Then, the maximum mean function value and the maximum knowledge gradient of the temporarily updated fused Gaussian process are evaluated.  These quantities can then be used to compute the value-gradient utility that would result if the sample, $\left(\mathbf{x} \,\,,\,\, f_i^q(\mathbf{x})\right)$, was realized from information source $i$.  This is given as
\begin{equation}
\label{utility2}
U_{\mathbf{x},i}^q = \mu_{\textrm{fused}}^{*,\textrm{temp}} +  \max_{\mathbf{x}' \in \chi} \,\,\, \nu^{KG}(\mathbf{x}').
\end{equation}
This process is repeated for all $N_q$ samples by removing the previously added sample and augmenting with the next new sample. The expected value-gradient utility obtained from adding alternative $\mathbf{x}$ to information source $i$ is then computed as
\begin{equation}
\label{avgKG}
EU_{\mathbf{x},i} = \frac{1}{N_q} \sum_{q = 1}^{N_q} U_{\mathbf{x},i}^q. 
\end{equation}
This expected utility is evaluated for all the alternatives and all the information sources. By denoting $C_{\mathbf{x},i}$ as the cost of querying information source $i$ at design $\mathbf{x}$, which is often computational expense for computational models, we find the query $(i_{N+1},\mathbf{x}_{N+1})$ that maximizes the expected value-gradient utility per unit cost, given by
\begin{equation}
(i_{N+1},\mathbf{x}_{N+1}) =\argmax_{i \in \{1, \dots, S\}\,\,,\,\, \mathbf{x} \in \mathbf{X}_{f}}  \frac{EU_{\mathbf{x},i}}{C_{\mathbf{x},i}}.
\label{KGC}
\end{equation}
After querying the design point $\mathbf{x}_{N+1}$ from the selected information source, $i_{N+1}$, the corresponding Gaussian process and afterward, the fused Gaussian process, are updated. This process repeats until a termination criterion, such as exhaustion of the querying budget, is met. Then, the optimum solution of Eq.~\ref{prob} is found based on the mean function of the fused Gaussian process.  This design is then to be created at ground truth.  Information from this creation can then be fed back into the framework if more resources are allocated. We note here that the computational complexity of the knowledge gradient policy is $\mathcal{O}(M^2 \log M)$, where $M$ is the number of alternatives considered~\cite{frazier2009knowledge}.  Thus, the computational complexity of the value-gradient querying policy is $\mathcal{O}([(S+1)M]^2 \log [S+1]M)$, where the $S+1$ terms represent each of the $S$ information sources and the fused information source.  We also note, that value-gradient policy inherits the capabilities of the knowledge gradient policy in terms of ability to handle nonlinear and multimodal objective functions.  




\section{Demonstration: Information Fusion}
\label{sec:demo_fusion}
In this section, we demonstrate the use of our multi-information source fusion approach to the dual phase steel application.  For this demonstration, we fuse information from the three physics-based reduced-order materials information sources with potentially non-uniformly sampled inputs.  We compare the results to ground truth data collected from the finite element RVE model. We consider three different cases.  The first case involves uniformly sampled data for each information source.  The second case involves non-uniformly sampled information from the information sources, with a large region where each information source is only sparsely sampled.  The third case involves non-uniform sampling of the information sources where each information source is sampled well over a small region of the input space and sparsely elsewhere.  In each case, the multi-information source fusion approach taken here performs well, and is far superior to using any of the information sources in isolation. We conclude this section with a novel analysis of the effective number of information sources used to make the fused estimate at each point in the domain.  This analysis provides a clear indication of our ability to exploit correlation for fusion, since without correlation information, only a single information source can confidently be used at a given point in the domain.

\subsection{Case 1: Uniform Sampling}
\begin{figure}[h!]
\begin{center}
\includegraphics[width=.4\textwidth]{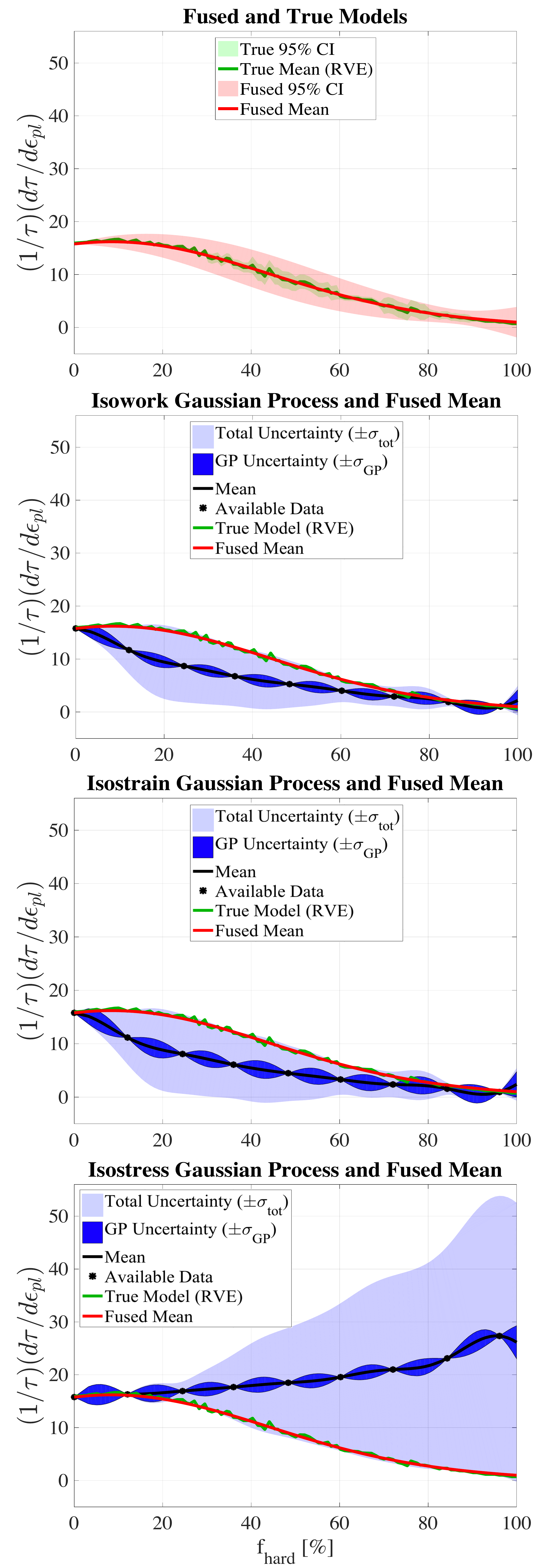}
\end{center}\vspace{-2ex}
\caption{The fused model and Gaussian processes of the isowork, isostrain and isostress models in comparison with the true (RVE) model.}
\label{results11} 
\end{figure}

In this case, the ground truth is assumed to have been sampled previously at nine uniformly spaced points in the input domain.  Each information source, that is, the isostrain, isostress, and isowork models has been evaluated at the points where ground truth information is known.  The nine sampled points for each information source are used to construct Gaussian process surrogate models for each.  These are shown as black lines through the nine black dots on the bottom three plots of Fig.~\ref{results11}.
The dark shaded region on each of these plots represents the uncertainty associated with each Gaussian process respectively.  The ground truth data was used to estimate the discrepancy of each information source from ground truth over the domain.  This additional uncertainty is the lighter shaded regions in the bottom three plots of the figure.  We note here again that we always assume the information sources are unbiased.  This assumption allows us to avoid simply fitting each information source to the ground truth data, which would result in eliminating useful information in each information source.  On each plot of Fig.~\ref{results11}, the ground truth is represented with the jagged green line and the result of our multi-information source fusion approach is represented by the smooth red line. 

From the isostrain, isowork, and isostress subplots, it is clear that no single information source performs well across the domain.  Indeed, over much of the domain each source performs poorly.  However, as can be seen in the upper left plot, our fused information source is an excellent match to ground truth.  This is further evidenced by the data provided in Tab.~\ref{MSE1}, where the mean squared errors (MSE): 
\begin{equation}
\textrm{MSE}_i = \frac{1}{N} \sum_{j=1}^{N} (g(\mathbf{x}_{j}) - f_i(\mathbf{x}_{j}))^2,
\end{equation}
and mean Kullback-Liebler divergences ($MD_{KL}$):
\begin{equation}
MD_{KL,i}(g \| f_i) = \frac{1}{N} \sum_{j=1}^N \int_{-\infty}^{\infty} p_g(\mathbf{x}_{j}) \log \frac{p_g(\mathbf{x}_j)}{p_{f_i}(\mathbf{x}_j)} \textrm{d}g,
\end{equation}
between the ground truth and each information source are presented.  Here, $p$ represents the probability density function of the information source given by the subscript.
\begin{table}[h!]
\caption{The mean squared errors (MSE) and the mean Kullback-Leibler divergences ($MD_{KL}$) between the true model (RVE) and the obtained models in Fig.~\ref{results11}.}
\begin{center}
\label{MSE1}
\renewcommand{\arraystretch}{1}
\begin{tabular}{c l l l}
\hline
Model & \;\; MSE & $MD_{KL}$ \\
\hline
Fused Model & \;\;\; 0.11 & 2.47 \\
Isowork Model & \; 13.29 & 5.56e+03  \\
Isostrain Model & \; 17.23 & 7.16e+03 \\
Isostress Model & 193.66 & 8.59e+04 \\
\hline
\end{tabular}
\end{center}
\vskip -.01cm
\end{table}
From the table, we see that the fused source is a significant improvement over any of the individual sources and can be used to reliably predict the ground truth over the whole domain.  

For this particular example, a subtle but crucial aspect of the use of the reification approach to information source fusion is revealed.  For the input region $f_{\textrm{hard}} \in (95,100]$, each information source overpredicts the truth.  However, as can be seen in the top plot of Fig.~\ref{results11}, the fused information has overcome the bias of each information source to match the ground truth nearly exactly. If the correlation between the information source discrepancies were not learned, this would not be possible.  This is due to the fact that the mean of a fused set of uncorrelated normal distributions is greater than the smallest mean and less than the largest mean in the set.  That is, the fused estimate can never overcome the bias that occurred here.  

\subsection{Case 2: Large Sparsely Sampled Region}
In this case, each information source has been sampled seven times, with all but one of the seven points being in the region $[0,50]$. These seven points are not necessarily the same for each information source.  We assume we have ground truth information at each of the seven points for each information source.  This situation could occur if, for example, different groups have available different sets of ground truth but are unaware of other data or are unwilling to share this information with other groups.  
\begin{figure}[h!]
\begin{center}
\includegraphics[width=.4\textwidth]{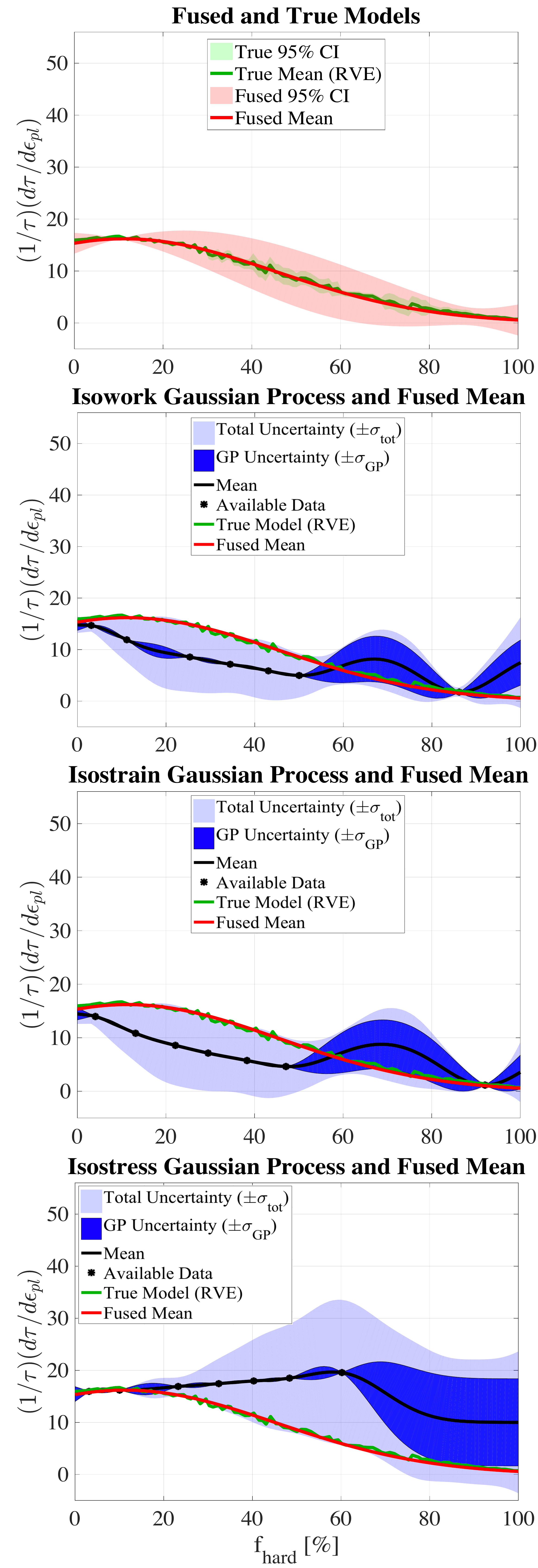}
\end{center}\vspace{-2ex}
\caption{The fused model and Gaussian processes of the isowork, isostrain and isostress models in comparison with the true (RVE) model when a few number of data is available in one region.}
\label{results12} 
\end{figure}
The key purpose of this demonstration case is to show the performance of our methodology over a poorly interrogated region of the domain when correlation information has been learned elsewhere over the domain.  The results of this demonstration case are shown in Fig.~\ref{results12}.
The information on each plot is presented in the same manner as that of Fig.~\ref{results11}.  As can be seen from the top plot, the fused information source again performs well, albeit with more predictive uncertainty than Case 1, which is to be expected. The MSE and $MD_{KL}$ values between the ground truth and each information source are given in Tab.~\ref{MSE2}.
\begin{table}[h!]
\caption{The mean squared errors (MSE) and the mean Kullback-Leibler divergences ($MD_{KL}$) between the true model (RVE) and the obtained models in Fig.~(\ref{results12}).}
\begin{center}
\label{MSE2}
\renewcommand{\arraystretch}{1}
\begin{tabular}{c l l l}
\hline
Model & MSE & $MD_{KL}$ \\
\hline
Fused Model & \; 0.18 & 4.51 \\
Isowork Model & 16.09 &  6.35e+03  \\
Isostrain Model & 18.85 &  9.72e+03 \\
Isostress Model & 65.85 &  1.70e+04 \\
\hline
\end{tabular}
\end{center}
\end{table}
\vskip -.01cm
Here, we see again that the fused information source is far superior to any information source in isolation.  We also see that the better sampled situation of the first demonstration case results in a more accurate fused information source. Of additional interest in this case is that the ability of our fusion approach to overcome bias of all information sources is more readily apparent.  Particularly, over the region $f_{\textrm{hard}} \in (60,85]$, all three information sources overpredict the ground truth.  However, the fused estimate has been pushed down towards the ground truth, away from the information source estimates.  We stress here that we consider each of the information sources as unbiased in their uncertainty quantification.  That is, the direction toward ground truth was not assumed by fitting each information source discrepancy to the ground truth in a biased fashion.  Indeed, there are no ground truth samples in this region.  The bias mitigation is due to the exploitation of correlations that have been learned through reification. 




\subsection{Case 3: Nearly Non-Overlapping Samples of Each Information Source}
\begin{figure}[h!]
\begin{center}
\includegraphics[width=.4\textwidth]{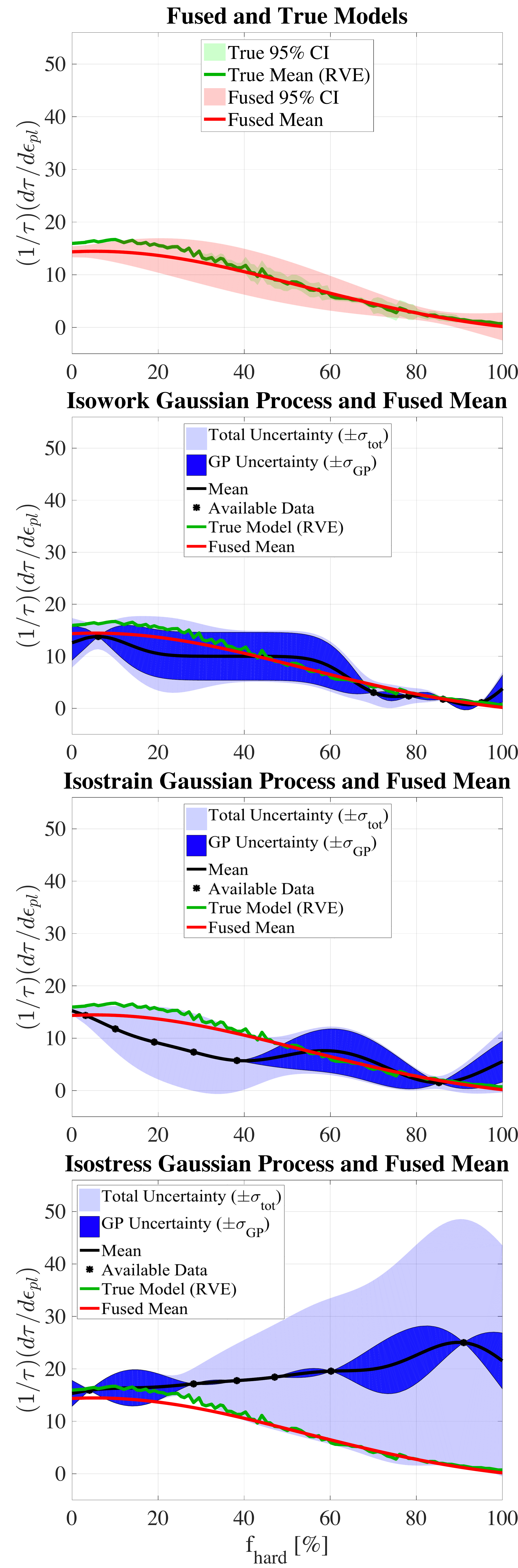}
\end{center}\vspace{-2ex}
\caption{The fused model and Gaussian processes of the isowork, isostrain and isostress models with data in different regions in comparison with the true (RVE) model.}
\label{results13} 
\end{figure}

In this case, each information source has been sampled a few times in a specific region of the domain.  The isostrain model was sampled generally in the left half of the domain, the isowork model was sampled generally in the right half of the domain, and the isostress model was sampled generally in the middle of the domain.  Ground truth was again used to quantify information source discrepancy but was not shared across information sources.  The key purpose of this demonstration case is to show the performance of our methodology when the information sources are essentially disparate in their knowledge of the quantity of interest over the domain.  The results of this demonstration case are shown in Fig.~\ref{results13}.
The information on each plot is presented in the same manner as that of Fig.~\ref{results11}. As can again be seen by the top plot, our approach performs well. The MSE and $MD_{KL}$ values between the ground truth and each information source are given in Tab.~\ref{MSE3}.
\begin{table}[h!]
\caption{The mean squared errors (MSE) and the Kullback-Leibler divergences ($MD_{KL}$) between the true model (RVE) and the obtained models in Fig.~(\ref{results13}).}
\begin{center}
\label{MSE3}
\renewcommand{\arraystretch}{1}
\begin{tabular}{c l l}
\hline
Model & MSE & $MD_{KL}$ \\
\hline
Fused Model &  \;\; 1.31 & 2.97 \\
Isowork Model &  \;\; 6.38 & 465.73  \\
Isostrain Model & \; 15.67 & 9.07e+03 \\
Isostress Model & 178.48 & 4.40e+04 \\
\hline
\end{tabular}
\end{center}
\end{table}


Of particular interest in this case is the performance of the fused information source in comparison to each individual source where that source was most heavily sampled.  We can see clearly from Fig.~\ref{results13} that the fused information source is a better approximation to ground truth in the left half of the domain than the isostrain model, where the isostrain model was most queried.  The same is true when comparing the middle of the domain estimates from the fused information source to the isostress results, and the right half of the domain results from the fused information source and the isowork model.  In each case, the fused information source is able to leverage the limited information from the other information sources to significantly outperform the information source that was heavily queried from in a given region.

\subsection{Effective Independent Information Sources}
To complete the demonstration of our multi-information source fusion approach, we define and present a novel number of effective independent information sources index.  The index measures the effective number of independent information sources partaking in the fused estimate at each point of the input domain. To define the index, we first consider the normalized change of variance that occurs when information sources are fused together at a given point.  This change can be written as
\begin{equation}
\frac{\Delta \sigma^2(\mathbf{x})}{\sigma_*^2(\mathbf{x})} = 1 - \frac{1}{\sigma_*^2(\mathbf{x}) \mathbf{e}^\top \tilde{\Sigma}(\mathbf{x})^{-1} \mathbf{e}},
\end{equation}
where $\Delta \sigma^2(\mathbf{x})$ is the variance reduction at $\mathbf{x}$ from the current best information source's variance, $\sigma^2_*(\mathbf{x})$, at that point.  Then, for any number, $S$, of {\em independent} information sources, each with variance $\sigma_*^2(\mathbf{x})$ at $\mathbf{x}$, we can write
\begin{equation}
\frac{\Delta \sigma^2(\mathbf{x})}{\sigma_*^2(\mathbf{x})} = 1 - \frac{1}{S}.
\end{equation}
Thus, the number of effective independent information sources with variance $\sigma_*^2(\mathbf{x})$ at the point $\mathbf{x}$ is given as
\begin{equation}
I_{\textrm{eff}} = \sigma_*^2(\mathbf{x}) \mathbf{e}^\top \tilde{\Sigma}(\mathbf{x})^{-1} \mathbf{e}.
\end{equation}
This index takes the value $S$ when there are $S$ independent sources with the same variance. If any sources have a larger variance, they will not contribute as much to the variance reduction at that point, and $I_{\textrm{eff}}$ will be less than $S$. Thus, effective independent information sources, as measured by this index, are relative to the best source at a given point.  We note here also that for highly correlated information sources, Eq.~\ref{e:WinklerVarianceGen} can result in variance decreases that are larger than would occur with independent information sources.  This generally occurs as a result of very similar but biased in the same direction information sources.  The ability to exploit this situation is a feature of the reification approach.  

\begin{figure}[h!]
\begin{center}
\includegraphics[width=.47\textwidth]{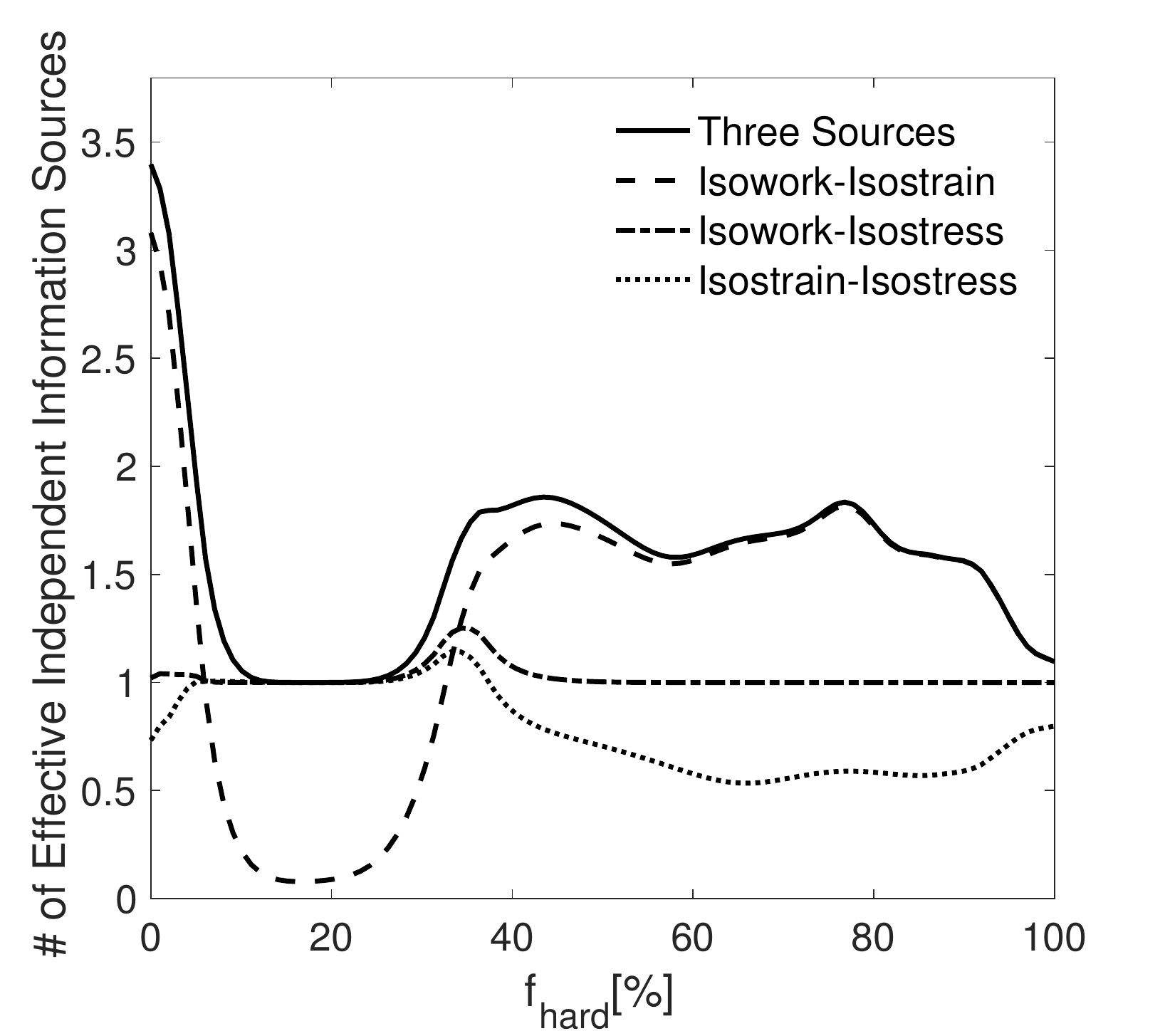} 
\end{center}\vspace{-2ex}
\caption{Number of effective independent information sources, $I_{\textrm{eff}}$ as a function of $f_{\textrm{hard}}$ for demonstration Case 1.}
\vskip -.3cm
\label{EffSources} 
\end{figure}

The effective independent information source index for demonstration Case 1 is shown in Fig.~\ref{EffSources}.
The figure includes the $I_{\textrm{eff}}$ for the three source case, as well as each pair of two sources.  The two source indices are still considered with respect to the best of the three information sources at a given point.  This leads to the possible situation where pairs of information sources are contributing less than one effective information source.  

For the fused approximation of Case 1, shown in the top plot of Fig.~\ref{results11}, it is interesting to note that the $I_{\textrm{eff}}$ is not large over the input space, as shown in Fig.~\ref{EffSources}.   For this particular problem, there is often one source that is much more uncertain than the other two at each point in the domain.  This renders that source's contribution to $I_{\textrm{eff}}$ to be very small.  For example, isostrain has large variance for low to medium values of $f_{\textrm{hard}}$ and isostress has large variance for large values. 

From the pairwise curves, it is clear that initially the isowork-isostrain pair is driving the fused approximation. It is also clear that in this region of low values of $f_{\textrm{hard}}$, the fusion process is exploiting the high correlation between these two sources and is performing better than three independent sources could.  The  isostress model takes over the approximation around $f_{\textrm{hard}} \approx 10\%$.  This holds until $f_{\textrm{hard}} \approx 30\%$, where all three sources are contributing to the prediction. At $f_{\textrm{hard}} \approx 40\%$, the isowork-isostrain pair again drive the prediction.  This continues until until the end of the domain is reached.  Thus, while all three information sources do not contribute equally over the domain, they are all three required to make the fused approximation presented in Fig.~\ref{results11}.

Though the effective independent information source analysis presented here relied only on the $I_{\textrm{eff}}$, which was derived through the discrepancy quantification and reification process, the analysis is consistent with the fundamentals of mechanics for these information sources and this application.  This provides evidence that such an analysis could be used to aid in the construction of a more sophisticated physics-based model from models using simplified assumptions.  That is, this index provides information about when certain assumptions are valid and when they are not.  The index also provides a means of valuing a new evaluation of an information source over the domain.  For example, this analysis reveals that sampling the isostress model on the interval $f_{\textrm{hard}} \in [40,100]$ will provide little value in terms of effective information sources when an isostrain and isostress model are also available.  Such a valuation could prove useful in a resource constrained process for estimating a quantity of interest with many possible information sources available.

\section{Demonstration: Multi-Information Source Optimization}
\label{sec:demo_optimization}

In this section, we demonstrate the application of our framework to the optimization of the ground truth strength normalized strain hardening rate for the dual-phase steel application.  We stress here that the purpose of our framework is the optimization of ground truth.  That is, our motivation is the creation of a myopic multi-information source optimization framework for addressing Eq.~\ref{prob} in the context of materials design.  Thus, we seek to identify the best candidate for a ground truth experiment with whatever resources we have available.  Once those resources are exhausted, a ground truth experiment takes place based on the recommendation of our framework.  The result of that experiment can then be fed back into the framework.  If more resources are then allocated, perhaps on the basis of promising results, then the framework can be employed again.

The specific demonstration consists of the use of the three reduced-order models (isostrain, isostress, and isowork) to query the impact of quantifiable microstructural attributes on the mechanical response of a composite microstructure---in this case a dual-phase steel. The \emph{ground truth} in this case is the finite element model of the dual-phase material.  The objective is the maximization of the 
(ground truth) normalized strain hardening rate at $\epsilon_{pl} = 1.5\%$. The design variable is the percentage of the hard phase, $f_{\textrm{hard}}$, in the dual-phase material.  We assume that our resources limit us to five total queries to (any of) the information sources before we must make a recommendation for a ground truth experiment.  Given promising ground truth results, five more queries can be allocated to the information sources.  The framework is initialized with one query from each information source and one query from the ground truth.  This information is used to construct the initial intermediate Gaussian process surrogates.

\begin{figure}[h!]
\begin{center}
\includegraphics[width=.45\textwidth]{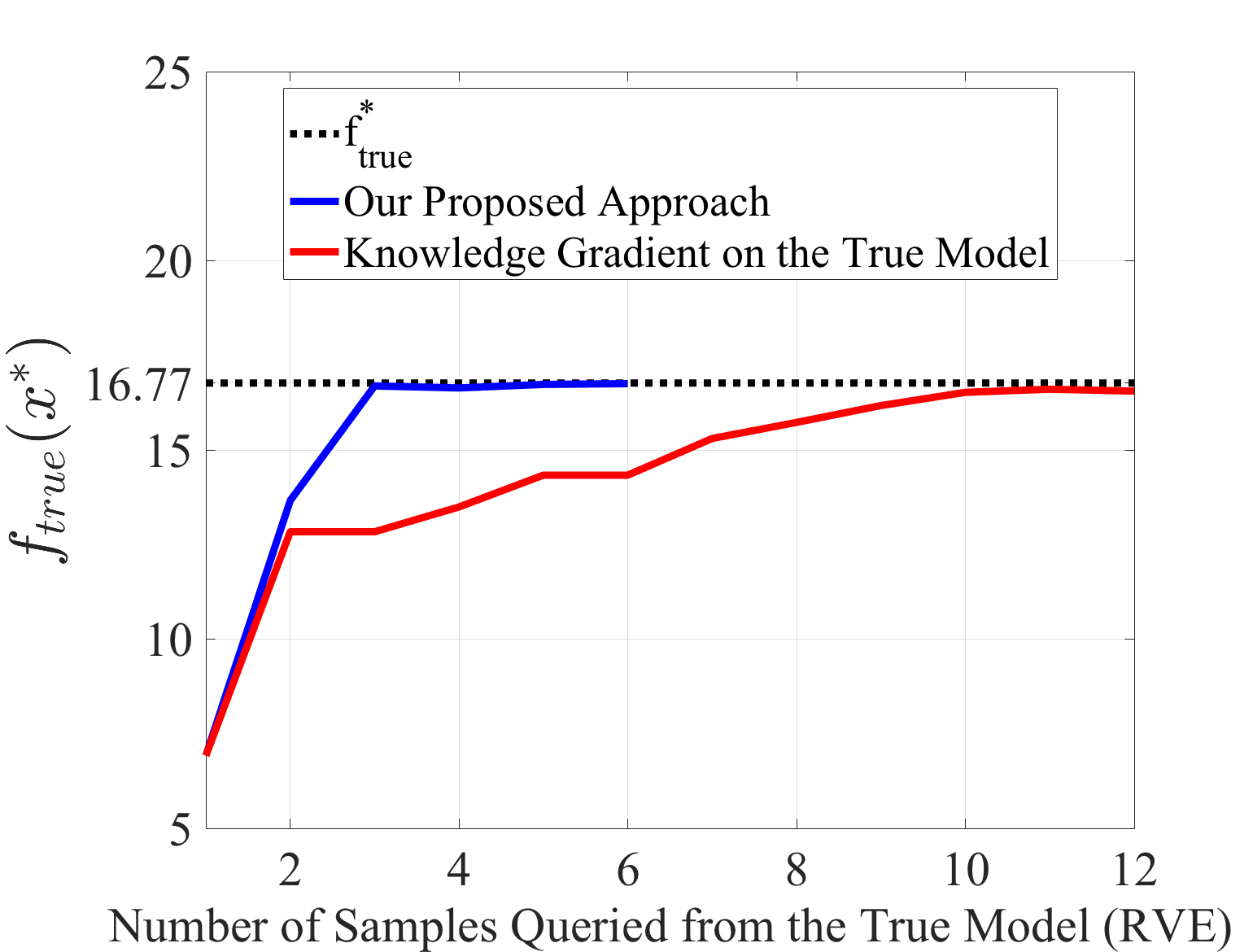}
\end{center}
\caption{The optimal solution obtained by our proposed approach and by applying the knowledge gradient on a GP of only the true data (RVE) for different number of samples queried from the true model.}
\vskip -.5cm
\label{convergence} 
\end{figure}

The value-gradient policy of our framework was used to select the next information source and the location of the query in the input space for each iteration of the process.  For comparison purposes, the KG policy operating directly on the ground truth was also used to reveal the gains that can be had by considering all available information sources. For this, a Gaussian process representation was created and updated after each query to ground truth. The convergence results of our proposed approach using all information sources and the KG policy on the ground truth are shown in Fig.~\ref{convergence}.
On the figure, the dashed line represents the optimal value of the ground truth quantity of interest.  It is clear from this figure that our approach outperformed the knowledge gradient applied to directly to ground truth, and in doing so, saved considerable expense by reducing the number of needed ground truth experiments.  
The superior performance of our approach can be attributed to its ability to efficiently utilize the information available from the three low fidelity information sources to better direct the querying at ground truth.
We note that the original sample from ground truth used for initialization was taken at $f_{\textrm{hard}} = 95\%$, which is far away from the true optimal.  This can be seen below in Fig.~\ref{iteration6} in the left column.  
\begin{figure*}[t]
\begin{center}
\includegraphics[width=1\textwidth]{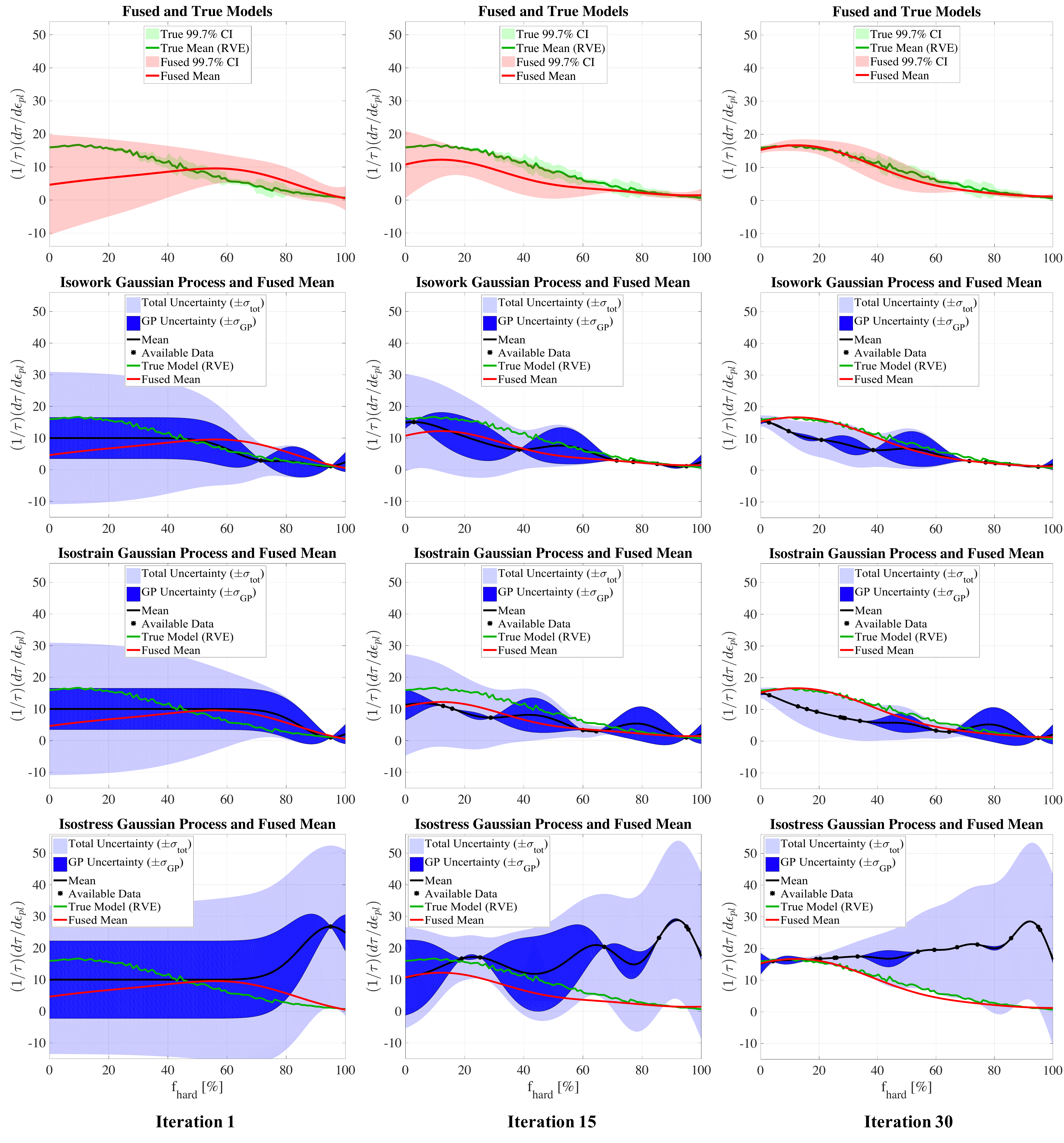}
\end{center}
\caption{The fused model and Gaussian processes of the isowork, isostrain and isostress models in comparison with the true (RVE) model in iterations 1, 15 and 30.}
\label{iteration6} 
\end{figure*}
Thus, the framework, by leveraging the three \emph{inexpensive} available information sources, was able to quickly direct the ground truth experiment to a higher quality region of the design space.

Table~\ref{opt1} presents the results of each ground truth experiment conducted according to the recommendation of our framework.
\begin{table}[h!]
\caption{The optimal solution obtained by the fused model, and the true value at the obtained optimum design point. The true optimal solution by the RVE model is $({\mathbf{x}}^* \, ,\, f^*) = (8.54 \,\, ,\,\, 16.77)$.}
\begin{center}
\label{opt1}
\vskip -.5cm
\begin{tabular}{c l l l l}
& \\ 
\hline
Experiment & ${\mathbf{x}}^*_{fused}$ & $f^*_{fused}$ & $f_{true}({\mathbf{x}}^*_{fused})$\\
\hline
2 & 28.64 & 7.50 & 13.66 \\
3 & 10.05 & 11.98 & 16.69 \\
4 & 10.55 & 13.92 & 16.64 \\
5 & 9.55  & 15.47 & 16.73 \\
6 & 8.80  & 16.71  & 16.75 \\
\hline
\end{tabular}
\end{center}
\vskip -.5cm
\end{table}
From the table we see that the third recommendation for a ground truth experiment produces a nearly optimal design.  
The final three experiments show that little more is gained in terms of ground truth objective and that the fused model has learned more about the ground truth in that region.  At this point, it is likely that more resources would not be allocated to this design problem and the framework was able to successfully find the best design.

Updates to each information source Gaussian process surrogate model and the fused model representing our knowledge of ground truth are also shown in Fig.~\ref{iteration6} for iterations 1, 15, and 30 of the information source querying process.  Here, an iteration occurs when an information source is queried.  This is distinct from any queries to ground truth.
As can be seen from the left column, the first experiment from ground truth and the first query from each information source told us little about the location of the true objective.  However, on iteration 15, the fused model, shown by the smooth red curve, has identified the best region of the design space, although it under-predicts the ground truth at this point.  We note that at this point, only 3 expensive ground truth experiments have been conducted.  By iteration 30, the fused model is very accurate in the region surrounding the optimal design for ground truth.  At this point, six ground truth experiments have been conducted.  From the figure, and also from Fig.~\ref{Figure: hardening}, it is clear that none of the information sources share the ground truth optimum.  The \emph{ability of the framework to find this optimum rested upon the use of correlation exploiting fusion}, and would not have been possible using traditional methods.

\begin{figure}[h!]
\begin{center}
\includegraphics[width=.5\textwidth]{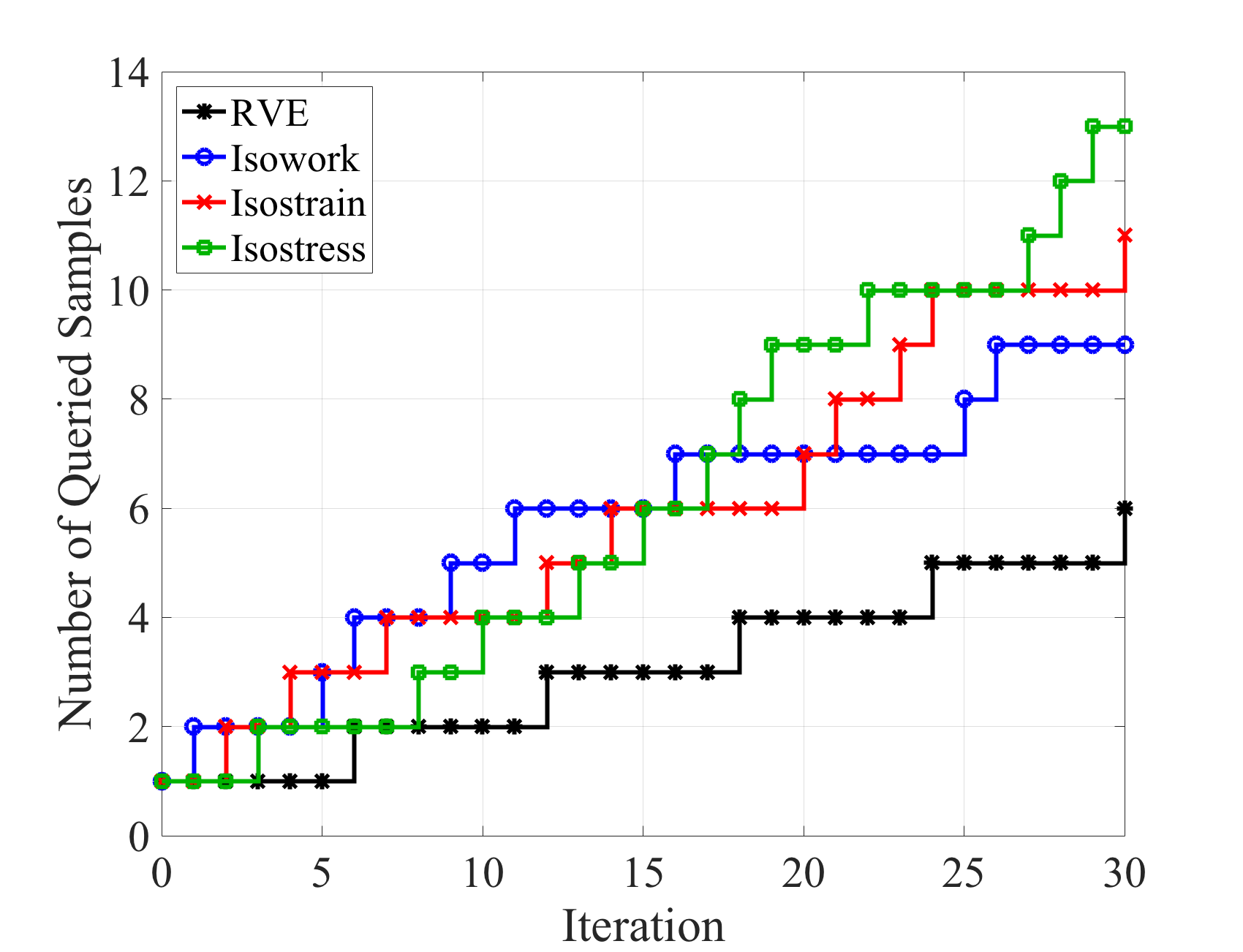}
\end{center}
\caption{Number of samples queried from the true model (RVE) and the information sources in each iteration.}
\label{funcount} 
\end{figure}

To conclude this demonstration we present the history of the queries to each information source and the ground truth.  This information is provided in Fig.~\ref{funcount}.
Note, that the iteration now counts queries to each information source as well as ground truth experiments.  From the figure, it is clear that all three information sources are exploited to find the ground truth optimal design, implying that, however imperfect, \emph{all} sources available to the designer must be used, in an optimal manner, in order to identify the optimal ground truth.


\section{Conclusions and Future Work}
\label{sec:conc}

In this paper we first presented and demonstrated a correlation-exploiting multi-information source fusion approach. The method included new extensions to the fusion of any number of correlated information sources, as well as the creation of a novel effective independent information source index.  The fusion methodology was demonstrated on microstructure-sensitive performance prediction for ductile dual-phase materials.  In all cases, the proposed fusion approach performed exceptionally well, far exceeding the predictive capabilities of any individual information source.  This provides evidence that our approach to information source fusion is highly applicable to the challenge of integration in ICME tools.  

We then presented and demonstrated a myopic multi-information source optimization framework.  The framework focused on determining the next information source to query and where in the input domain to query it by trading off resource expense and gains expected in ground truth objective function quality.  To value each next potential query we presented a novel value-gradient policy, which seeks to maximize a two-step lookahead utility based on immediate value and the knowledge gradient for a potential next step.  The framework was demonstrated on the optimization of ground truth strength normalized strain hardening rate for a dual-phase material.  The results of the demonstration revealed the promise of this framework as a suitable methodology for answering the MGI call for accelerating the materials development cycle.  

In the near-term, the information fusion framework developed here will be validated against larger sets of ground truth data and be demonstrated in higher dimensions.  The framework will also be extended to handle information sources with misaligned input-output interfaces, which is a key challenge facing the ICME community.

Moreover, the optimization framework will be extended to handle multiple objectives and studied for scalability to high dimensional input spaces. Additionally, we will explore the possibility of carrying out optimal sequential queries in which the sources of information are not input/output aligned. A specific scenario, for example, would be combining sources that establish relationships between processing parameters/conditions and microstructure with sources that connect microstructures to properties/performance. Much remains to be done, but this work presents a plausible research program towards the realization of the promise of ICME, which in the end rests on tool (or information source) \emph{integration}.

\begin{acknowledgment}
The authors would like to acknowledge the support of the National Science Foundation through grant No.\ NSF-CMMI-1663130, {\em DEMS: Multi-Information Source Value of Information Based Design of Multiphase Structural Materials}. Arroyave would also like to acknowledge the support of the National Science Foundation through grant No.\ NSF-CMMI-1534534, \emph{DMREF: Accelerating the Development of Phase-Transforming Heterogeneous Materials: Application to High Temperature Shape Memory Alloys}. Allaire and Arroyave would also like to acknowledge the support of the National Science Foundation through grant No.\ NSF-DGE-1545403, \emph{NRT-DESE: Data-Enabled Discovery and Design of Energy Materials (D$^3$EM)}. 

%
\end{acknowledgment}

%
% ...

\bibliographystyle{asmems4}




\end{document}